%% file: resubmit3_ms.tex
\documentclass[12pt,preprint]{aastex}
\usepackage{natbib}

\def\eg{\mbox{e.g.}}
\def\ie{\mbox{i.e.}}
\def\kms{\mbox{km~s$^{\rm -1}$}}

\newcommand{\HI}{H\,{\sc i}}

\slugcomment{}

\shorttitle{The Magellanic Stream and Debris Clouds}

\shortauthors{B.-Q. For, L. Staveley-Smith, D. Matthews, N.~M. McClure-Griffiths}

\begin{document}

\title{The Magellanic Stream and Debris Clouds}

\author{B.-Q. For\altaffilmark{1,2}, 
L. Staveley-Smith\altaffilmark{1},
D. Matthews\altaffilmark{3},  
N.~M. McClure-Griffiths\altaffilmark{4}}

\affil{$^1$International Centre for Radio Astronomy Research, University of Western Australia, 
35 Stirling Hwy, Crawley, WA, 6009, Australia; biqing.for@icrar.org}
\altaffiltext{2}{SIEF John Stocker Fellow}
\affil{$^3$Centre for Materials and Surface Science, La Trobe University, 
Melbourne, VIC, 3086, Australia} 
\affil{$^4$CSIRO Astronomy and Space Science, Epping, NSW, 1710, Australia} 

\begin{abstract}
We present a study of the discrete clouds and filaments in the Magellanic Stream
using a new high-resolution survey of neutral hydrogen (\HI) conducted with H75
array of the Australia Telescope Compact Array, complemented by single-dish data
from the Parkes Galactic All-Sky Survey (GASS). From the individual and 
combined datasets, 
we have compiled a catalog of 251 clouds and list their basic
parameters, including a morphological description useful for identifying cloud
interactions. We find an unexpectedly large number of head-tail clouds in the
region. The implication for the formation mechanism and evolution is discussed.
The filaments appear to originate entirely from the Small Magellanic Cloud and 
extend into the northern end of the Magellanic Bridge. 
\end{abstract}


\section{INTRODUCTION\label{intro}}

The Magellanic Clouds (MCs) are the closest extragalactic neighbors to our Galaxy. 
Their gaseous component was discovered in atomic hydrogen as a coherent 
stream originating from the MCs, namely the Magellanic Stream 
(MS; \citealp{WW72,Mathewson74}) and Leading Arm (LA; \citealp{Putman98}). 
The MS is trailing the MCs and has been subdivided into six main concentrations 
(MS~I--VI) \citep{Mathewson74}. 
Although this subdivision is an oversimplification, 
it is widely used to describe different regions along the stream 
in the literature. 
 
Recent detailed studies around the outskirt of the MS have made new discoveries.  
Several new filaments that are parallel with the MS \citep{Westmeier08} and 
extended near the northern tip of the MS \citep{Nidever10} have been found. 
These suggest that the MS and LA now spans a total of $\sim200\degr$ in 
length across the sky and is $\sim30\degr$ wide at its widest point.  
The LA, a counterpart of the MS, is leading the \HI\ gas stream 
and is morphologically very different from the MS. It is dominated by 
three distinctive high-velocity cloud (HVC) complexes (LA I--III; 
\citealp{Putman98,Bruns05}), 
and LA IV which is formed by a population of HVCs as 
defined by For et al.~(2013; hereafter FSM13), 
or as a single HVC complex defined by \citet{Venzmer12}.  

The formation and origin of the MS and LA are still under debate. 
There are two well-studied physical mechanisms that may be the cause 
for the formation of the MS and LA: ram-pressure stripping of gas from the MCs due 
to interaction with the Galactic halo (\eg, \citealp{MD94,Mastropietro05}), 
and tidal interaction between the Milky Way and the MCs (\eg, \citealp{GN96,Connors06}).
Both scenarios were modeled and required orbits 
to reproduce the global observed \HI\ column density and 
velocity distribution of the MS and LA.  
High-precision Hubble Space Telescope (HST) proper motion measurements of 
the MCs \citep{K06b,K06a,K13} favor 
either an unbound orbit for the MCs with 
a first passage scenario \citep{Besla07} 
or an eccentric long period orbit \citep{SL09}. 
The proposed new scenarios based on the HST proper motion measurements 
draw controversy as tidal and ram-pressure 
stripping require the MCs to be relatively close to the Milky Way for a  
long interaction time (multiple orbits).
However, recent simulations have shown that tidal force of the LMC acting on 
the SMC alone can create the MS and LA before the MCs were accreted by the Milky Way 
in the first passage scenario \citep{Besla12,DB12}.
Simulations with a bound orbit and extra drag 
have also successfully reproduced the observed bifurcation 
of the MS and remain consistent with the second epoch proper motion 
data \citep{DB11}. A blowout hypothesis has been proposed by \citet{Nidever08}. 
They suggest that the supergiant shells in the southeast HI overdensity 
region of the LMC are blown out to larger radii where ram-pressure and/or 
tidal forces can more easily strip the gas into the MS and LA. 
Connections between gas in the LMC and the Bridge, 
Stream and Leading Arm have previously been pointed by 
other authors \citep{McGee86,Putman98,Putman03,LSS03}, 
though not as the main source of gas feeding the stream. 

An observational approach 
to trace the origin of the MS and LA uses metallicity measurements 
as an indicator. Early studies of metallicity 
measurements from the HST spectra of background sources toward the MS and LA II  
constrain their origin to the MCs ($Z=0.2-0.4$ solar; \citealp{Lu94,Lu98,Gibson00}). 
Subsequent studies with the HST and/or FUSE spectra suggest that the MS and LA 
originate from the SMC \citep{Sembach01,Fox10}. 
Recent metallicity measurements based on the VLT/UVES and HST/COS 
spectra of four background active galactic nuclei strongly support a scenario of which 
most of the stream gas were stripped from the SMC \citep{Fox13}. However, 
a subsequent study of the same series reports a much higher metallicity (S/H =0.5 solar) 
in the inner Stream toward background quasar, Fairall 9 \citep{Richter13}. 
The overall chemical abundances of MS toward Fairall 9 are significantly different 
as compared to other sightlines suggest a complex enrichment history of the stream. 
They favor the explanation of local $\alpha$-enrichment by massive stars followed by separation 
from the MCs before nitrogen enrichment occurs. 
They also state that the finding supports the dual-origin as identified by \citet{Nidever08}     
but is not a requirement.
Nevertheless, there is still a short-coming in the ability of 
theoretical models to reproduce the 
general observed LA features and complex filamentary structure of the MS. 
Physical properties of HVCs, such as the multiphase structures formed via 
hydrodynamical interactions with the hot halo, are 
seldom able to be considered in simulations. 

Large-scale \HI\ surveys have provided huge amounts of information for studying 
the gaseous structures surrounding our Milky Way. The most notable surveys are the 
\HI\ Parkes All-Sky Survey (HIPASS; \citealp{Barnes01}), the Leiden-Argentine-Bonn 
all-sky \HI\ survey (LAB; \citealp{Kalberla05}) and the Galactic 
All-Sky Survey (GASS; \citealp{MG09}) in the Southern hemisphere.
HIPASS is an extragalactic survey with a coarse velocity resolution ($\Delta v \sim18$~\kms) 
but is very sensitive and covers a large range in velocity. The data were employed to create an HVC catalog 
(\citealp{Putman02}, hereafter P02) and to recover the extended structure 
of the Magellanic System \citep{Putman03}. 
GASS is a Galactic survey with a much better  
spectral resolution ($\Delta v=1$~\kms) and far more suitable to study the gaseous feature 
of the Magellanic System. LAB also has good spectral resolution ($\Delta v \sim1.3$~\kms) 
but less angular resolution than GASS. The LAB data were employed by \citet{Nidever08} 
for their study on the Magellanic System.
Other smaller scale studies for the Magellanic System or subregions 
include, for example, the narrow-band Parkes \HI\ survey \citep{Bruns05}, 
\HI\ Galactic studies with Arecibo $L$-band Feed Array (GALFA-HI; \citealp{Stanimirovic08}) 
and a Green Bank Telescope survey \citep{Nidever10}.

In this paper we study the overall morphology of the MS using interferometric 
and single dish data, the general distribution, the physical properties and morphological 
properties of its HVCs. The aims are: (1) to understand 
the formation and origin of the MS, (2) to investigate 
the effect of interaction with the Galactic halo, and 
(3) to provide constraints for simulations based on the observed 
properties. 
In \S2 and \S3, we describe the observations and 
the procedures for source finding. We present the catalog 
and cloud properties in \S4 and \S5. 
Cloud morphology, kinematic distribution and interpretation of 
the distributions are presented in \S6. We discuss 
the implications of the HVCs on the formation of the MS 
by comparing the observed properties with theoretical models in 
\S7. Summary and conclusions are given in \S8.

\section{OBSERVATIONS AND DATA\label{red}}

A detailed description of the observing strategy and data reduction 
for the ATCA high-resolution Magellanic Stream survey is given in \citet{Matthews14}. 
A description of GASS is given by \citet{MG09} and \citet{Kalberla10}. 
A brief summary of the observations and characteristics of the data is given below.

The ATCA high-resolution Magellanic Stream survey covers a 500 deg$^{2}$ field of the 
MS using the H75 configuration of the ATCA. MS~I to MS IV, 
part of Small Magellanic Cloud (SMC) and the Interface Region (IFR) are covered in this 
survey. The observations were carried out 
over a period from 2005 to 2006, which resulted in $\sim$180 hours of total 
observing time. 
The entire area was divided into 33 regions with 154 pointing centers per region, 
resulting in 5082 pointing centers. 
Each pointing center was separated by 20$\arcmin$, arranged in a hexagonal grid, 
observed for 20~s and revisited 6 times during an average of 10 hour observing. 
The resulting ATCA data have an angular resolution of $413\arcsec\times330\arcsec$, 
a brightness sensitivity of 210~mK and a velocity resolution of 1.65~\kms\ after Hanning smoothing. 
The survey covers Local Standard of Rest velocity ($V_{\rm LSR}$) 
between $-315$ and +393~\kms.

GASS is an \HI\ survey of the entire 
sky south of declination $+1\degr$ using the 20~cm multibeam receiver 
on the Parkes radio telescope. 
This survey covers $V_{\rm LSR}$ 
between $-$400 and +500~\kms. 
The second data release\footnote{http://www.astro.uni-bonn.de/hisurvey/gass} 
employed in this study has been corrected for 
stray radiation and radio frequency interference, 
has a channel width of 0.82~\kms, a spectral resolution 
of 1~\kms, a brightness temperature ($T_{\rm B}$) rms sensitivity of 57~mK and 
an angular resolution of 16$\arcmin$. 

The reduced ATCA images are grouped into 11 data cubes (see 
\citealp{Matthews14} for the boundary of each data cube).  
To obtain the large-scale structure information missing from 
the interferometric data, \citet{Matthews14} combined the ATCA data with 
the GASS data, hereafter H75GASS. 
The GASS data were matched to the angular dimensions of each ATCA data cube prior merging. 
The data were merged in the image domain as detailed by \citet{Matthews14}.
The merged data have an average $T_{\rm B}$ sensitivity of 250~mK, and are in 
equatorial coordinates.

To precisely and conveniently describe the position of 
various gaseous features along the MS, 
a Magellanic coordinate system has been used by many. 
For example, \citet{Wakker01} defined the north pole of the Magellanic coordinate 
system as Galactic longitude ($\l$) 180\degr\ and latitude ($b$) 0\degr. 
Its equator passes through the Galactic equator at $l=90\degr$ and $b=270\degr$ and passes 
through the south Galactic pole. 
However, this definition does not place the equator exactly along the MS. 
A new coordinate system, Magellanic Stream coordinate, 
has since been introduced by \citet{Nidever08} 
to better represent the projection of the MS along the equator. 
This new coordinate system defines $L_{\rm MS}=0\degr$ at 
the center of the Large Magellanic Cloud (LMC), 
\ie, ($l,b$) = 280.47\degr, $-32.75$\degr\ \citep{vdm02}, 
and the pole at ($l,b$)=188.5\degr,$-7.5\degr$. The $L_{\rm MS}$ decreases 
in value toward the tip of the MS and increases in 
value toward the LA. We adopt this coordinate system for the rest of the paper.

\section{SOURCE FINDING\label{sf}}

\subsection{Preparation\label{prep}}

We converted the H75GASS data cubes into Galactic coordinates, 
with each covering a different range of velocities. 
To minimize the size of the data cubes, 
we created 12 subcubes covering the velocity range that 
contains obvious features of the MS. Any feature in the range $-$30~\kms\ $< V_{\rm LSR} <$ +18~\kms\ 
is indistinguishable from the strong Galactic \HI\ emission
and hence was excluded. 
Mild contamination from the Galactic \HI\ emission 
occurs at $-40\lesssim V_{LSR} \lesssim-30$~\kms\ and 
$+18\lesssim V_{LSR} \lesssim+30$~\kms, which predominantly affects region $-55\degr$ to $-40\degr$ 
in $L_{\rm ms}$. A mosaicked cube containing features 
of the MS at positive and negative LSR velocity was also created. 
A summary of velocity ranges for the original cubes and created 
subcubes is given in Table~\ref{cubesvel} 
(see \citealp{Matthews14} for the boundary of each data cube).   

For the GASS data alone, we also extracted data cubes that covered a similar region 
and velocity range, giving a total of four data cubes (negative velocity and positive velocity 
for both H75GASS and GASS) for source finding. 
Integrated \HI\ column density maps of the mosaicked H75GASS and GASS data are 
shown in Zenithal Equal Area (ZEA) projection in Figures~\ref{Hmom0} and \ref{Gmom0}. 

\subsection{Analysis\label{duchamp}}

We performed the source finding using {\it Duchamp}\footnote{Available at http://www.atnf.csiro.au/computing/software/duchamp/} 
version 1.2.2, a 3-dimensional source finder developed by \citet{Whiting12}. 
It is designed to search data cubes, 
merge detections and measure basic parameters of the detected sources. It also implements 
optional noise reduction routines, such as the wavelet reconstruction 
and spatial or spectral smoothing, to enhance the detectability of fainter sources.

Each of the input cubes was reconstructed with $\grave{a}$ {\it trous} wavelets 
to remove random noise prior to the search. 
A fixed threshold was specified for each run, and various thresholds 
were tested for each input cube to find the best source finding parameters.
Sources were detected only if they extended across a minimum of 10 spatial pixels and $\sim5$~\kms\ 
in velocity space (\ie, 5 and 3 velocity channels for GASS and H75GASS, respectively). 
Subsequently, the detected sources were compared to earlier detections and either merged 
with neighboring sources or added to the list as a single source. 

Despite the application of noise 
reduction technique, diffuse emission was better detected with the GASS data alone. 
Thus, we decided to use the H75GASS and 
GASS data independently to detect compact and diffuse sources, respectively. 
A fixed threshold of $\sim2\sigma$ above the background noise (57~mK) was employed for 
GASS positive and negative $V_{\rm LSR}$ cubes. 
We used fixed thresholds of $\sim3\sigma$ above the background noise (250~mK) 
for the H75GASS negative and positive $V_{\rm LSR}$ cubes.

\section{CATALOG\label{cat}}

A compilation of sources with basic parameters is presented in Table~\ref{catalog}. 
The catalog includes the following entries: the source identification number in Column 1; 
the designation with a prefix of HVC for high-velocity clouds and GLX for galaxies 
followed by the $l$, $b$ and central 
$V_{LSR}$ fitted with single Gaussian component in column 2; 
MS longitude ($L_{\rm MS}$) and MS latitude ($B_{\rm MS}$)
in columns 3--4; central $V_{LSR}$ in column 5; 
the central velocity in the Galactic 
Standard of Rest reference frame ($V_{GSR}$), 
defined by $V_{GSR}=220$~cos~$b$~sin~$l$~$+V_{LSR}$; 
the velocity in the Local Group Standard of Rest reference frame  ($V_{LGSR}$), 
defined by $V_{LGSR}=V_{GSR}-62$~cos~$l$~cos~$b$~$+40$~sin~$l$~cos~$b$~$-35$~sin~$b$ \citep{BB99} 
in columns 6--7; 
velocity FWHM, measured at 50$\%$ of peak flux in column 8; 
integrated flux, peak $T_{\rm B}$ and 
peak \HI\ column density ($N_{\rm HI}$) in columns 9--11;
semi-major axis, semi-minor axis and position angle in columns 12--14;  
warning flag, morphological classification, data origin and comment in columns 15--18. 

We detected a total of 574 sources in our initial search. 
In the case of the same source being detected in both GASS and H75GASS cubes, 
we examined the spectrum and the adopted the parameters of the source 
with higher spectral signal-to-noise ratio.
We also examined individual spectra and the data cube in order to eliminate false detections. 
Such detections are caused by background artifacts and noise peaks, which generally have a 
narrow line width.
The positions of all final sources 
were examined using the NASA/IPAC Extragalactic Database 
to identify galaxies. 75 sources overlap with the 
Galactic emission channels and/or lie at the spatial edge of the image. 
These sources are included in the catalog but their physical parameters are omitted. 
The final catalog includes a total of 251 HVCs and 2 galaxies (NGC~300 and WLM). 
NGC~55 and NGC~7733 are positioned near the MS in the sky but are not detected because 
they do not fall into the field-of-view of ATCA survey. NGC 55 is shown in 
the extracted GASS data cube but lies on an artifact.

The {\it Duchamp} software derived most of the parameters listed above 
except angular sizes and peak \HI\ column density. 
We adopted the same approach as described in FSM13 to determine 
the angular sizes and peak \HI\ column density. 
Two-dimensional Gaussian fitting was used to derive position angle, 
semi-major and semi-minor axes. It uses the brightness centroid for the fit, 
though can be inaccurate for distorted sources (\eg, clouds with two bright structures). 
Caution should be applied when interpreting the derived position angle. 
The peak $N_{\rm HI}$ was determined by locating the brightest pixel in the integrated 
\HI\ column density map of each source. 

In \citet{Westmeier12}, tests were performed to evaluate the reliability of 
{\it Duchamp} for parametrizing sources. Artificial unresolved 
and extended \HI\ sources were generated and tested with various parameters. 
They demonstrated that {\it Duchamp} is a powerful source finder with 
the capability of detecting sources down to low signal-to-noise ratio. However, 
parameters such as the integrated flux ($F_{\rm int}$) 
measured by {\it Duchamp} suffer from systematic errors. 
The integrated flux of faint sources is underestimated by {\it Duchamp}. 
In order to correct for this systematic error, we employed the same procedures
as described in FSM13 to derive the correction factor for sources in 
our catalog. In brief, we used the same stand-alone parametrization algorithm as 
employed in FSM13 to measure the integrated flux of our sources ($F^{'}_{\rm int}$). 
Subsequently, the ratio of measured $F^{'}_{\rm int}$ to $F_{\rm int}$ as a function of 
$F_{\rm int}$ in various bins was fitted with a polynomial. 
Figure~\ref{iflux_corr} shows the comparison 
between the functions derived in this study (blue dashed line) and FSM13 (red dotted line).
Both fitted functions represent the underestimation factor for a given $F_{\rm int}$ 
and behave similarly.
Either function can be used for correcting 
integrated fluxes measured by $Duchamp$ in future work. 
In this paper, we adopted the function derived from sources in this catalog. 

As for derived peak $T_{\rm B}$ and
peak $N_{\rm HI}$ measured by $Duchamp$, 
corrections are not necessary, but 
small systematic errors of the order of 5--10$\%$ of the derived values are 
expected (FSM13). The velocity FWHM values measured by $Duchamp$ 
are generally accurate (see Figure~8 of \citealp{Westmeier12}).

The completeness of our catalog depends on the detection rate 
of {\it Duchamp}. Simulations to check the detection rate were performed by 
FSM13. Fake clouds of various input parameters were injected into the 
data cube at random locations and searched by {\it Duchamp}. 
They find that {\it Duchamp} 
is generally reliable for detecting clouds with narrow lines, but likely 
to miss clouds with faint ($T_{\rm B}$=0.14--1.0~K) or broad velocity lines (velocity FWHM=16--30~\kms). 
The recovery rate for faint or broad velocity lines clouds is 80$\%$.
We expect the same detection rate for the present catalog.  
 
\section{Cloud Properties \label{properties}}

In Figure~\ref{dist}, we show histograms of 
MS longitude and MS latitude in the new coordinate system, 
$V_{\rm GSR}$, $V_{\rm LSR}$, peak \HI\ column density 
in logarithmic scale and velocity FWHM (from top left to bottom right).  
Galaxies and objects with a warning flag in the catalog are excluded in these histograms. 
We find that the number of HVCs decreases linearly (with coefficient of determination of 0.96) 
from $-20\degr$ to $-70\degr$ in $L_{\rm MS}$. 
The low number of HVCs at $L_{\rm MS}=-10\degr$ is due to the incomplete 
sky coverage. 
For $B_{\rm MS}$, the distribution is Gaussian with the peak at 0\degr. 
HVCs agglomerate in a narrow $B_{\rm MS}$ range of $\pm10\degr$.
We also find a flat distribution in $V_{\rm GSR}$ for the GASS data. Most HVCs 
found in the H75GASS data are distributed between 0 and 100~\kms\ in the 
GSR velocity frame. 
A bimodal distribution is seen in $V_{\rm LSR}$, with median of velocities 
$-$140 and +150~\kms. 
The lack of clouds in the negative $V_{\rm LSR}$ range 
is due to selection effects. 
The peak $N_{\rm HI}$ distribution is distinctly different between 
HVCs found in GASS and H75GASS data. Both exhibit a normal Gaussian distribution with 
the majority of low and high \HI\ column density clouds being detected 
in GASS and H75GASS data, respectively. The lower \HI\ column 
density limit in the two datasets is due to sensitivity. The gap
between the histograms appears to a resolution effect -- peak column densities will 
always be higher for the higher angular resolution H75GASS data.
Although not shown here, the distribution of 
peak $N_{\rm HI}$ is uniform with respect to $L_{\rm MS}$. The distribution
in average column density (integrated over $B_{\rm MS}$) is also uniform in the 
range $L_{\rm MS}=-20\degr$ to $-70\degr$. However, as shown in 
Figure 10 of \citet{Nidever10}, there is an exponential decrease thereafter.
The FWHM velocity distribution 
suggests that clouds in this region have a median value of 25~\kms\ and 
their overall distribution is 
similar to that found in the LA region (FSM13), 
except that there are no HVCs with a velocity FWHM greater than 80~\kms\ in the MS region. 

\subsection{Comparison with the P02 Catalog \label{comp}}

The P02 HVC catalog is based on reprocessed HIPASS data, which recovers 
slightly more extended emission than the original HIPASS processing 
(see \citealp{Putman03} for details). The search for objects was performed using 
a friends-of-friends cloud search algorithm \citep{Heij02}.   
The P02 catalog covers the entire right ascension range, 
the declination range of $\delta<$+2\degr\ and the velocity range of 
$-500 \leq V_{\rm LSR} \leq +500$~\kms. 
Since the LSR velocity range of $\pm90$~\kms\ does not exclude all the Galactic emission, 
an additional constraint based on the deviation 
velocity ($V_{\rm dev} > 60$~\kms), defined by \citet{Wakker91} 
to be the difference from a simple Galactic rotation model, was 
added by P02 to the selection criteria for their HVCs. 
They noted that the excluded 
objects may contain real HVCs with some having an appearance similar to the 
small-scale structure in the gaseous disk of the Milky Way.

In Figure~\ref{comparisons}, we show 253 sources in our catalog 
and 183 sources in P02 catalog that fall within our searched velocity and spatial 
volume. There are $\sim71$ sources identified as being the same source 
in two catalogs, with the majority of them detected in the GASS data (see right panel). 
This suggests that most of the sources detected in the H75GASS data are unique. 
Comparing the rates for identifying the same P02 sources between this study and FSM13, 
we find them to be $\sim40\%$ and $\sim50\%$, 
respectively.
Given the high brightness sensitivity of HIPASS as compared to GASS and H75GASS, 
P02 recovered many faint HVCs that were not detected in our dataset 
(\eg, the positive MS latitudes in MS~IV) in Figure~\ref{comparisons}.  
However, there is a lack of P02 detections in the IFR due to the nature of the 
reprocessed HIPASS data, in which emission is filtered out in individual 
channels in this region. 
The manner in which clouds are merged or broken up can affect overlap between the catalogs.
Both GASS and H75GASS have excellent spectral resolution, 
which allows us to resolve HVCs with narrow lines. 
The high-resolution H75GASS data recover a significant number of compact sources, 
which cannot be resolved in the lower resolution GASS and HIPASS data. 
This is clearly seen in the peak $N_{\rm HI}$ 
distribution (Figure~\ref{dist}). 
We show the velocity FWHM distribution of both catalogs in Figure~\ref{wvel_comp}. 
HVCs that overlap with the Galactic emission and/or lie at the spatial edge of the image 
and galaxies have been excluded in this plot. This figure shows that the excellent 
spectral resolution of GASS and H75GASS allows clouds with narrow-line components to be resolved. 
Most of the HVCs in our and the P02 catalog have velocity FWHM of 
15--25~\kms\ and  
35--45~\kms, respectively.  
While few in number, our catalog does contain HVCs with 
large velocity FWHM ($>40$~\kms). 

\section{\HI\ GAS AND KINEMATIC DISTRIBUTION\label{gaskin}}

The Magellanic Stream has traditionally been viewed as consisting of six main 
concentrations 
(\eg, \citealp{Mathewson74}). Subsequent observations with higher sensitivity 
revealed its complexity (see \eg, \citealp{Cohen82,Putman03,Bruns05,Nidever08}). 
Two distinct filaments were recovered, which run alongside each other 
with no clear starting point but appeared to be connected to the IFR 
\citep{Bruns05}. \citet{Bruns05} defined a separation between the IFR 
and start of the MS near $l=300\degr$ and $b=-61\degr$. 
\citet{Nidever08} studied the MS and LA using a Gaussian decomposition method 
on the LAB data and claimed to trace one of the MS filaments back to 
the southeast \HI\ overdensity region of the LMC.
In this section, we use both the H75GASS and GASS data (complementary to each other) 
to revisit the overall morphology and kinematics of the MS, and to discuss 
the detailed morphology of filaments and HVCs in the MS.  

\subsection{The overall morphology\label{overall}}

\begin{figure}
\epsscale{0.9}
\plotone{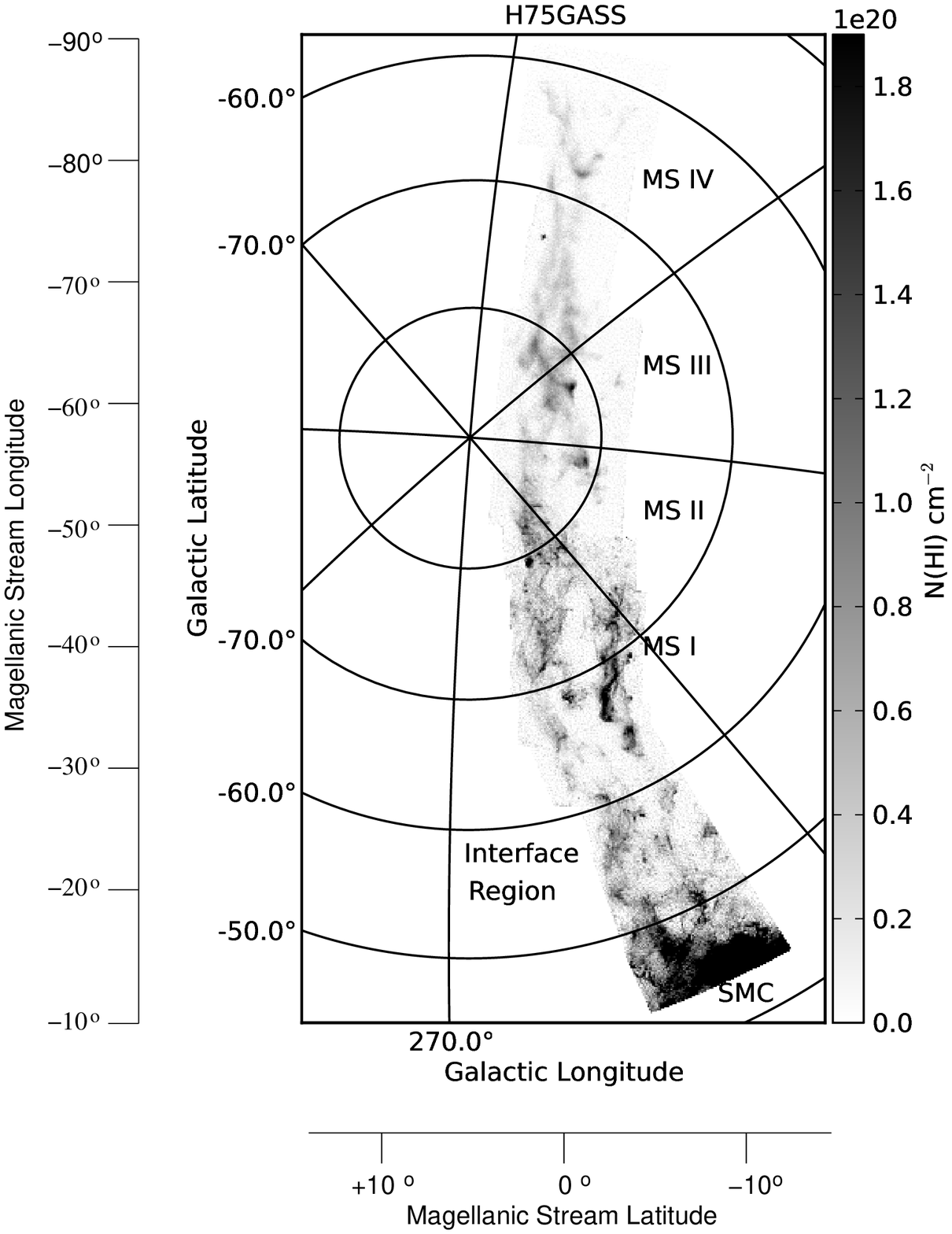}
\caption{Integrated \HI\ column density map of H75GASS for positive and negative velocities. 
The \HI\ column density ranges from 0 to $1.9\times10^{20}$~cm$^{-2}$. The locations 
of the Small Magellanic Cloud, the Interface Region and MS~I--IV are labeled. \label{Hmom0}}
\end{figure}

\begin{figure}
\epsscale{0.7}
\plotone{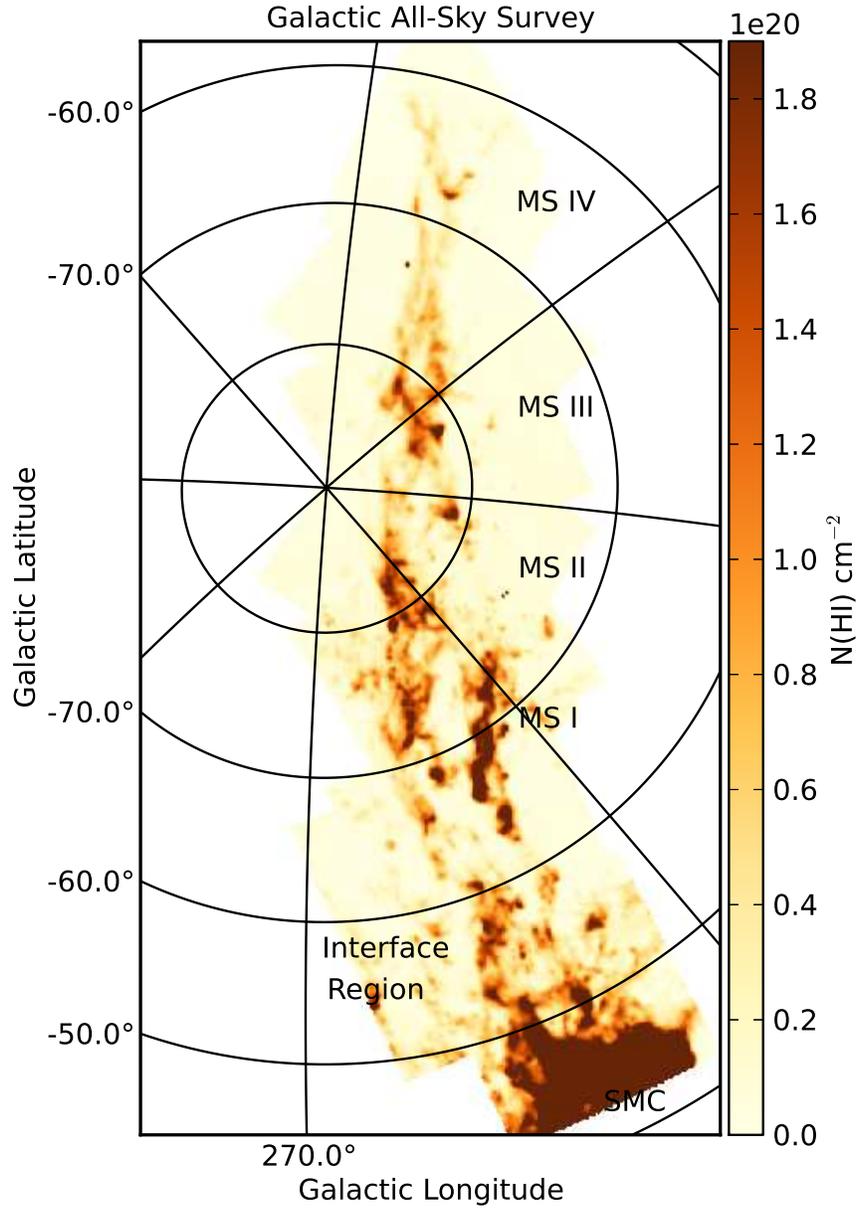}
\caption{Same as Figure~\ref{Hmom0} but for the 
GASS data only. A slightly larger region 
than H75GASS is used in this study. A different color scheme is used to 
distinguish the two data sets.\label{Gmom0}}
\end{figure}

\begin{figure}
\begin{center}
\includegraphics[angle=-90,scale=0.5]{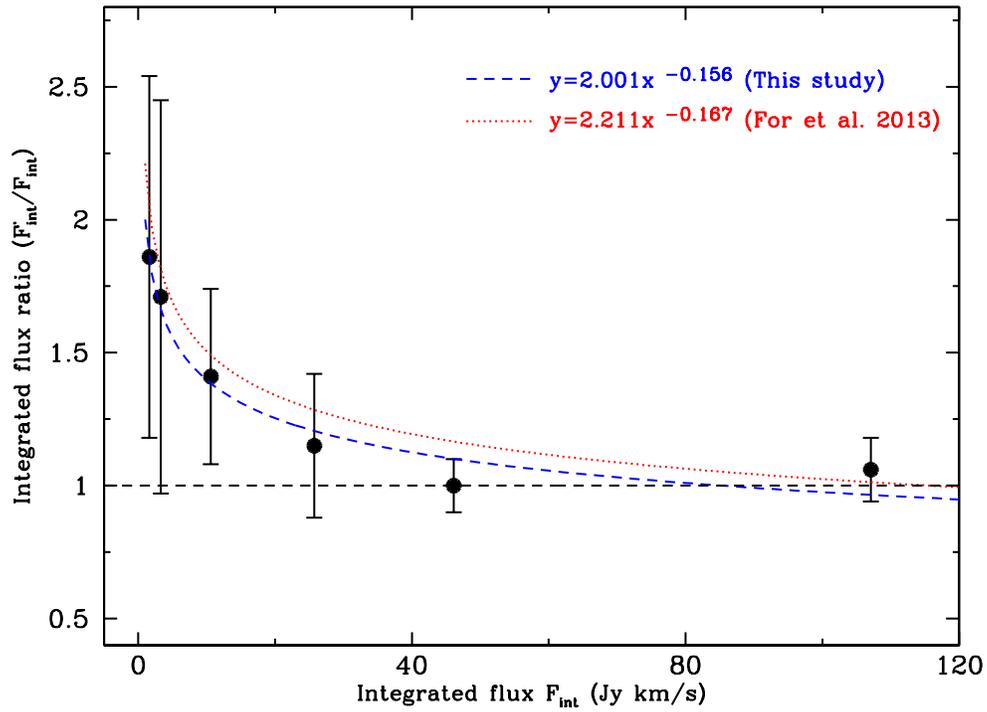}
\caption{Ratio of true integrated flux ($F^{'}_{\rm int}$) to 
integrated flux ($F_{\rm int}$) measured by 
$Duchamp$ as a function of $F_{\rm int}$. 
The red dotted and blue dashed lines are the fitting functions derived in 
\citet{For13} and this study, respectively. They represent an estimate 
of the correction factor to be applied to $Duchamp$ fluxes. 
\label{iflux_corr}}
\end{center}
\end{figure}

\begin{figure}
\epsscale{0.9}
\plotone{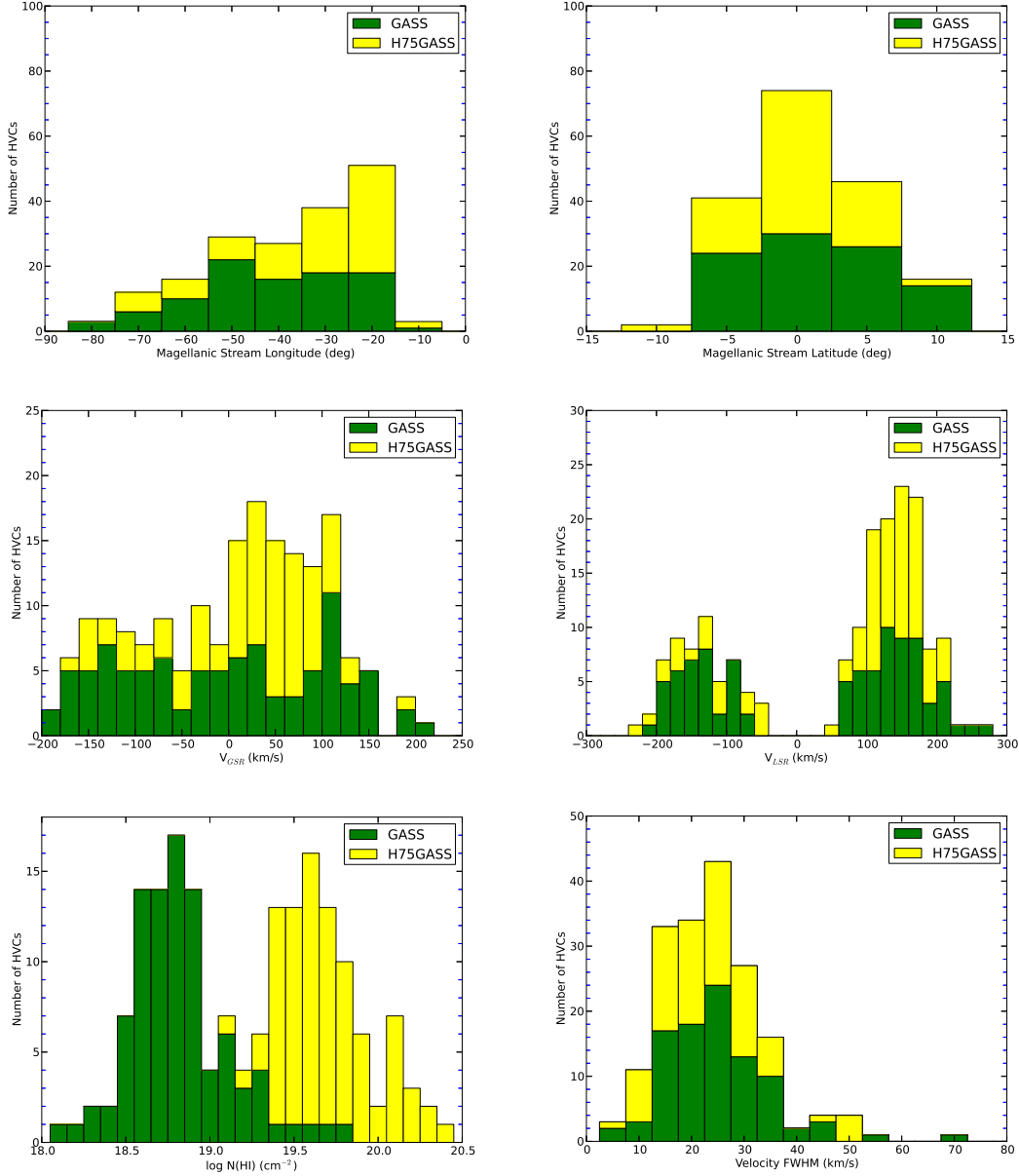}
\caption{Stacked histograms of $L_{\rm MS}$, $B_{\rm MS}$, $V_{\rm GSR}$, 
$V_{\rm LSR}$, peak \HI\ column density on a logarithm scale and 
velocity FWHM of HVCs identified in GASS (green) and H75GASS (yellow), 
from top left to bottom right, respectively. Excluded from the plots 
are galaxies and 75 HVCs that either overlap with the Galactic 
emission channels and/or lie at the spatial edge of 
the image. \label{dist}}
\end{figure}

\begin{figure}
\epsscale{1.0}
\plottwo{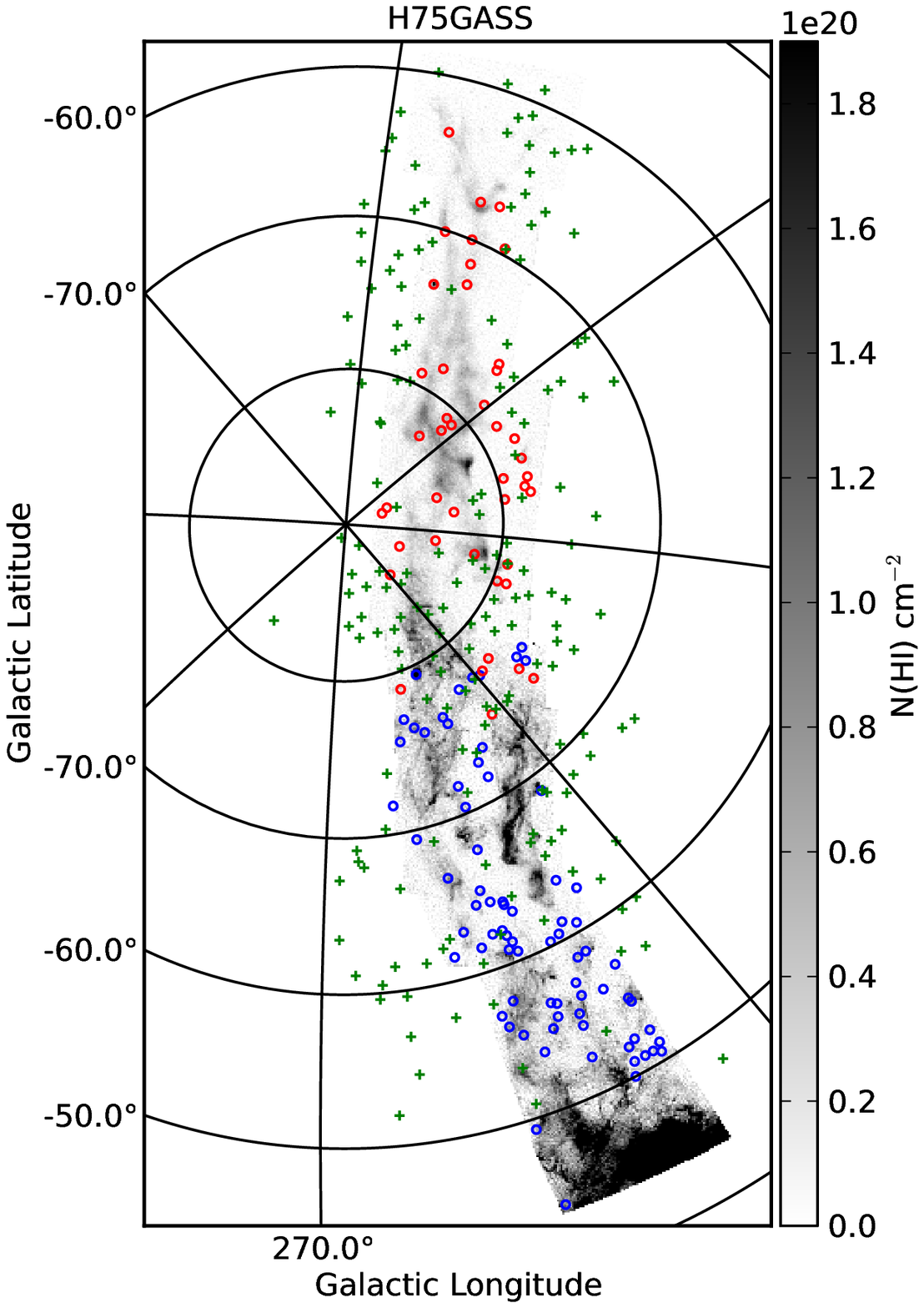}{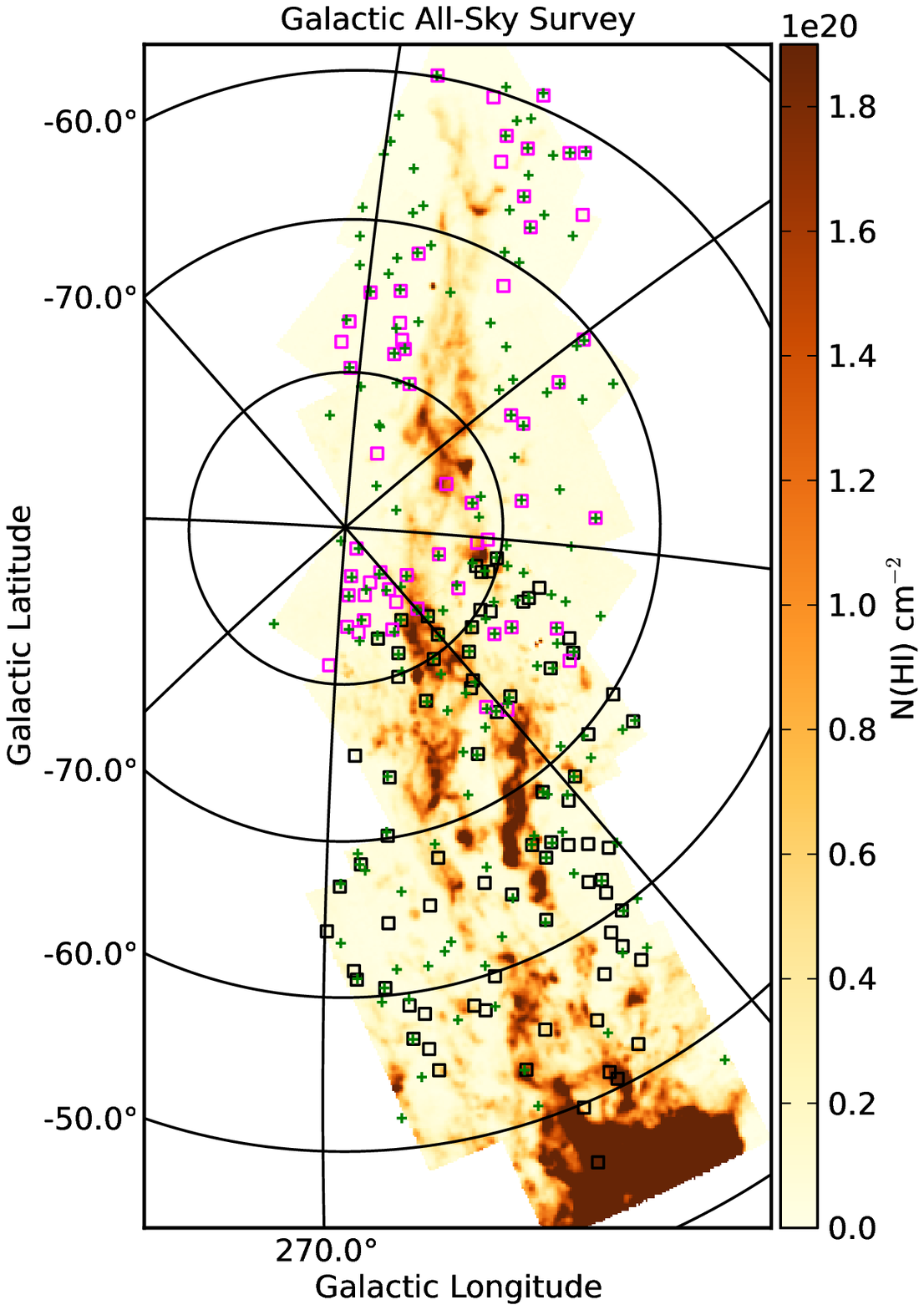}
\caption{{\it Left}: 
All 118 sources detected by $Duchamp$ with the H75GASS data 
(red and blue circles) and 183 sources detected by \citet{Putman02} 
(green pluses) are overlaid onto the integrated \HI\ column density map of 
H75GASS as shown in Figure~\ref{Hmom0}. 
The blue and red circles represent $Duchamp$ detections 
at positive and negative LSR velocities, respectively. 
{\it Right}: Same as left figure, except 135 sources detected by 
$Duchamp$ are from GASS data and are overlaid onto the 
\HI\ column density map of GASS as shown in Figure~\ref{Gmom0}. 
The black and magenta squares represent $Duchamp$ detections at 
positive and negative LSR velocities, respectively. 
The sources from \citet{Putman02} are within the same $V_{\rm LSR}$ 
range and the same region on the sky as our GASS data 
presented in this paper. 
\label{comparisons}}
\end{figure}

\begin{figure}
\epsscale{1.0}
\plotone{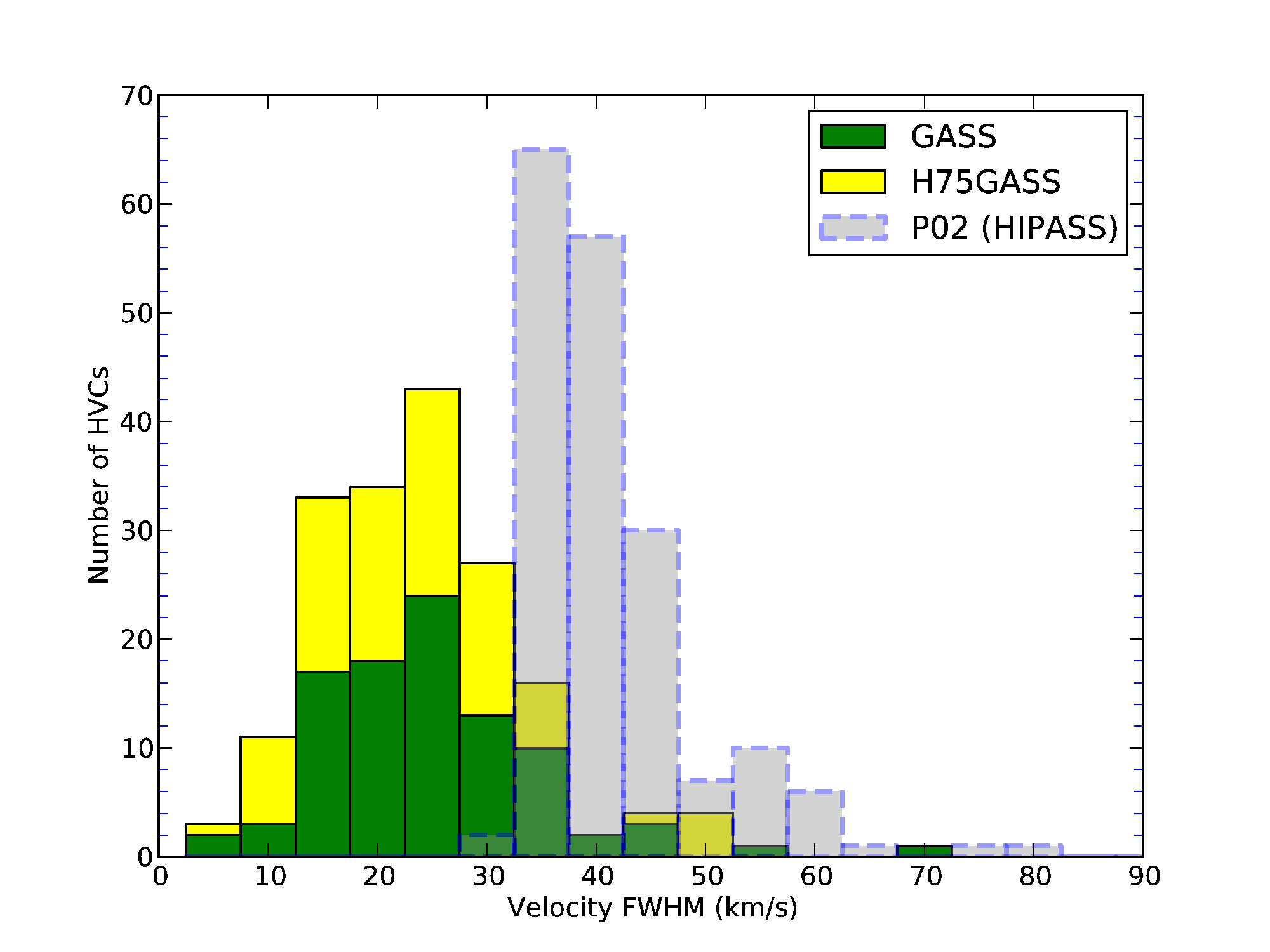}
\caption{Histograms of velocity FWHM of HVCs in this study 
(green and yellow) and P02 (grey).  
Excluded from the plots 
are galaxies and 75 HVCs that either overlap with the Galactic 
emission channels and/or lie at the spatial edge of the image.
Clouds with narrow lines  are recovered in this study 
thanks to the higher spectral resolution of GASS and H75GASS 
as compared to HIPASS. 
\label{wvel_comp}}
\end{figure}

In Figures~\ref{Hmom0} and \ref{Gmom0}, we show 
the integrated \HI\ column density maps of 
H75GASS and GASS data with nomenclature for various parts of the MS. 
In contrast to the study by \cite{Bruns05}, 
careful examination reveals that the two filaments of the MS are actually extended into the 
IFR, this redefining 
the starting point of the MS. 
This continuity is also observed in the LAB data studied by \citet{Nidever08}. 
They showed that both filaments are running parallel at the head of the stream. 
 
\begin{figure}
\epsscale{1.0}
\plotone{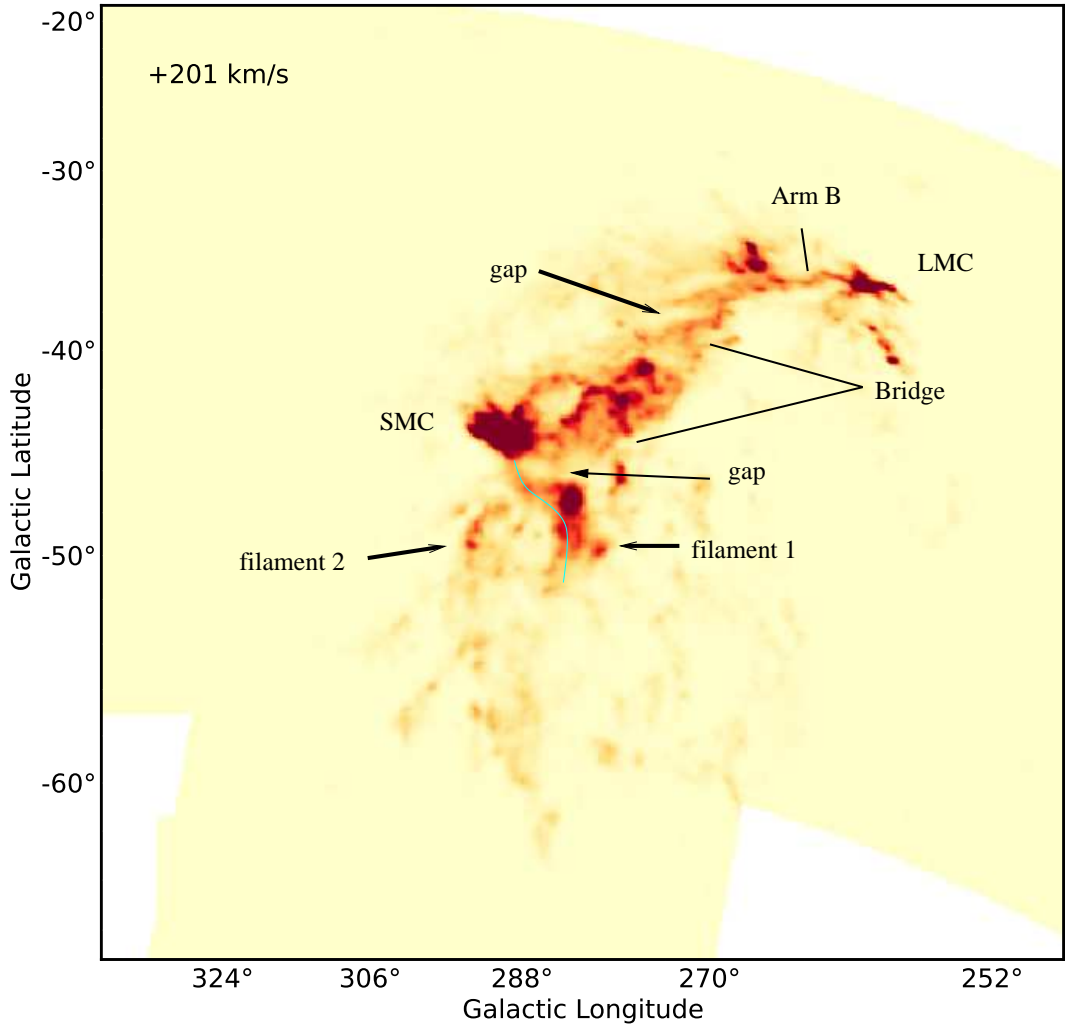}
\caption{Channel map at +201~\kms. The Small Magellanic Cloud, Large Magellanic Cloud, 
Magellanic Bridge, the dual-filaments at the head of the MS 
and Arm B are labeled. 
The gaps show filament 1 is not connected to the LMC. The cyan line marks the position of 
filament 1 which appears to be connected to the SMC both spatially and kinematically. 
\label{GASS_slice}}
\end{figure}

\begin{figure}
\epsscale{0.7}
\plotone{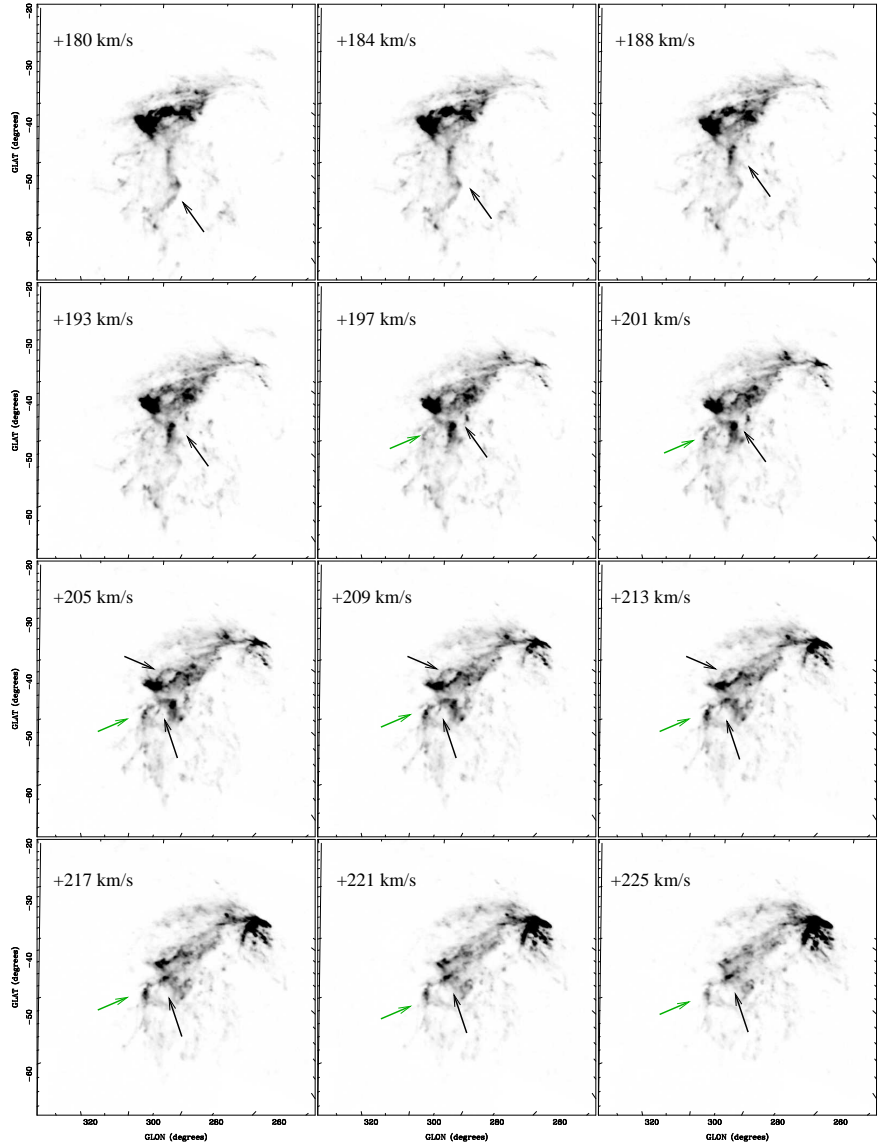}
\caption{Channel maps of GASS data cube from LSR velocity 
of +180~\kms\ to +225~\kms. These maps cover the area where 
Small Magellanic Cloud, Large Magellanic Cloud and 
the head of the MS are located. Black and green arrows indicate 
filament 1 and 2, respectively. \label{chan_map1}}
\end{figure}

\begin{figure}
\epsscale{0.7}
\plotone{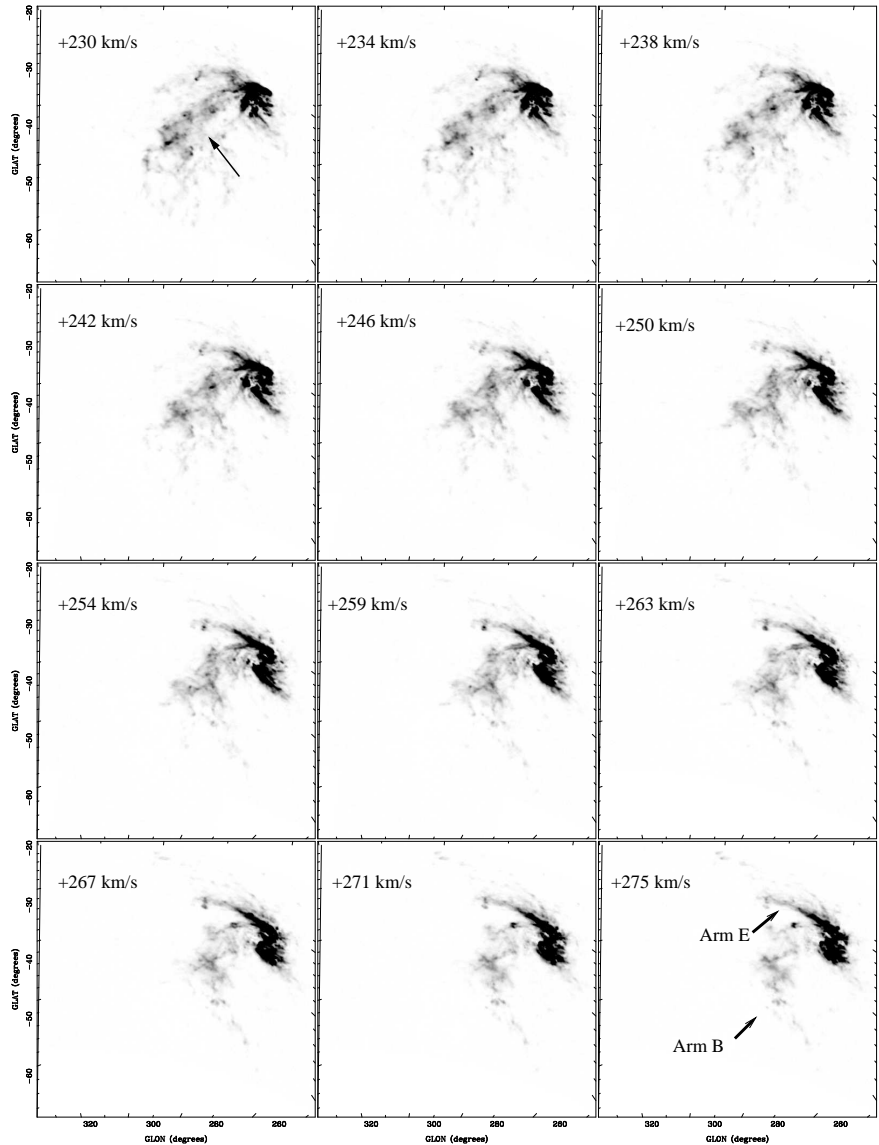}
\caption{Same as Figure~\ref{chan_map1}, except showing channel maps with 
LSR velocity from +230~\kms\ to +263~\kms. This velocity range is dominated 
by the \HI\ gas from the LMC and Magellanic Bridge. Black arrow in +230\kms\ 
channel map indicates the location of the Magellanic Bridge where filament 1 is now 
mixed in. Arm E and B are 
labeled in the last channel map and a sinusoidal pattern is seen. \label{chan_map2}}
\end{figure}

To trace the two filaments at the head of the stream, we use the GASS data, which has 
a better angular resolution than the LAB data. Figure~\ref{GASS_slice} 
shows the area of investigation, which covers the SMC, LMC and the 
head of the MS. The dual-filaments are also labeled as filament 1 and 2. 
Series of channel maps 
in the $V_{\rm LSR}$ range from +180 to +275~\kms, 
with an interval of $\sim$4~\kms, are shown in 
Figures~\ref{chan_map1} and \ref{chan_map2}. 
Tracing the most prominent filament 
(namely filament 1, indicated by black arrow) from 
velocity channels between +180 and +201~\kms, we find that it 
extends toward the SMC. A connection to the SMC emerges 
at +201~\kms\ and extends into the the northern end of the 
Magellanic Bridge (second black arrow as shown in +205~\kms\ channel map), 
where \HI\ gas from the SMC is dominant. The exact ending point of 
filament 1 is ambigious as it eventually mixes in with the Magellanic Bridge emission. 
In Figure~\ref{GASS_slice}, we show the channel map at +201~\kms in which distinct gaps 
are indicated. Gaps are common along the MS. 
Examining the channel maps beyond +255~\kms, we find that 
the gas from the LMC is dominant and eventually a distinct sinusoidal pattern emerges from 
the Arm B of LMC (see the last channel map of Figure~\ref{chan_map2}). 
This gaseous feature appears isolated and does not seem to connect to 
the filament 1. 
The exact starting point of filament 2 is hard to trace but 
the overall structure appears to be connected to the Magellanic 
Bridge through SMC (indicated by green arrow).

These two filaments appear to be widely separated at the head of the stream and become 
thinner and have decreasing \HI\ column density from the head to tail of the stream. 
They eventually turn into a clumpy filamentary structure, which is not shown 
in this study but revealed in the studies by \citet{Stanimirovic08} and \citet{Nidever10}.  
They also appear to be twisted several times along the stream, 
in a double-helical structure \citep{Putman03}. 

\begin{figure}
\epsscale{1.0}
\plottwo{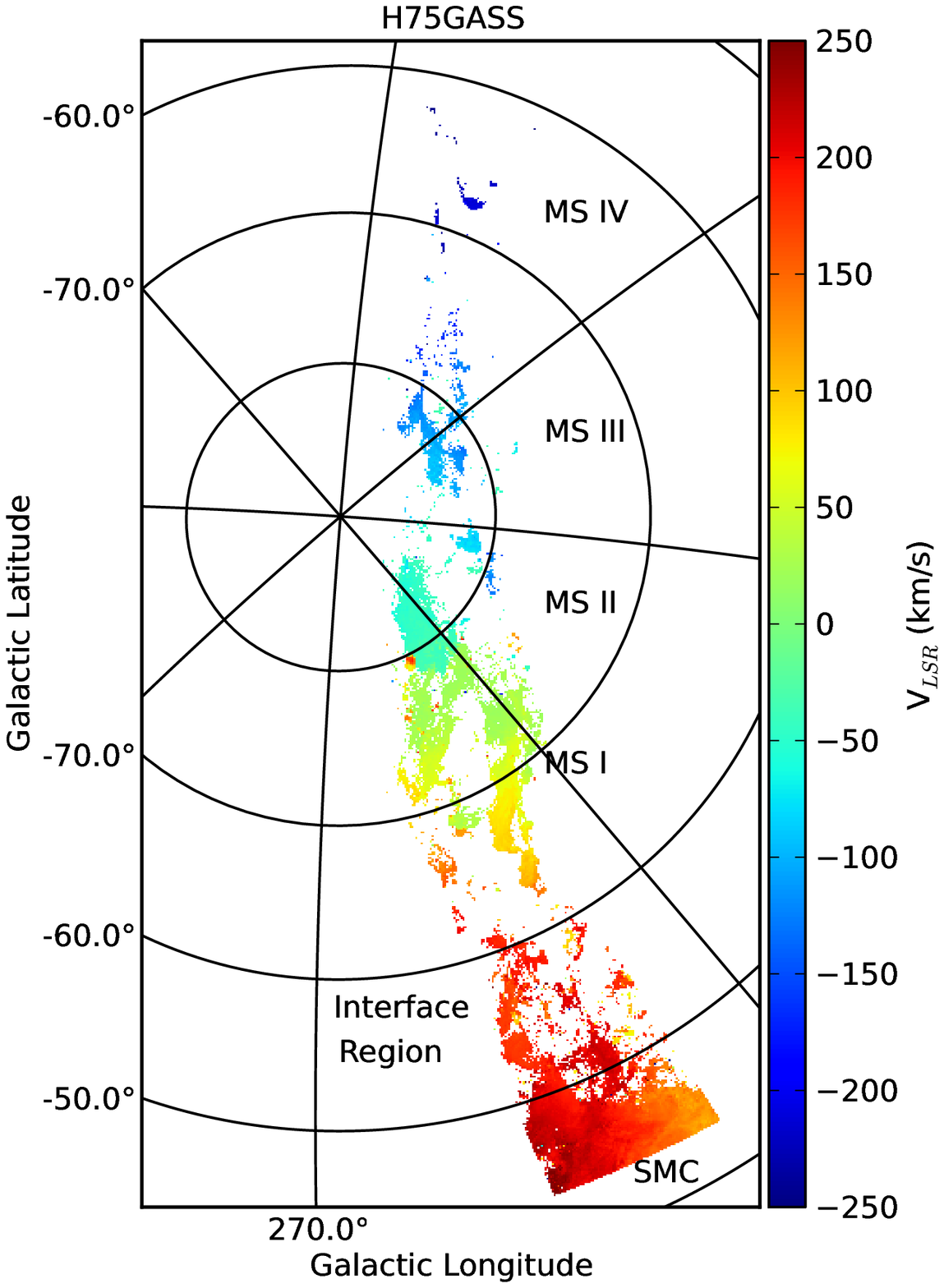}{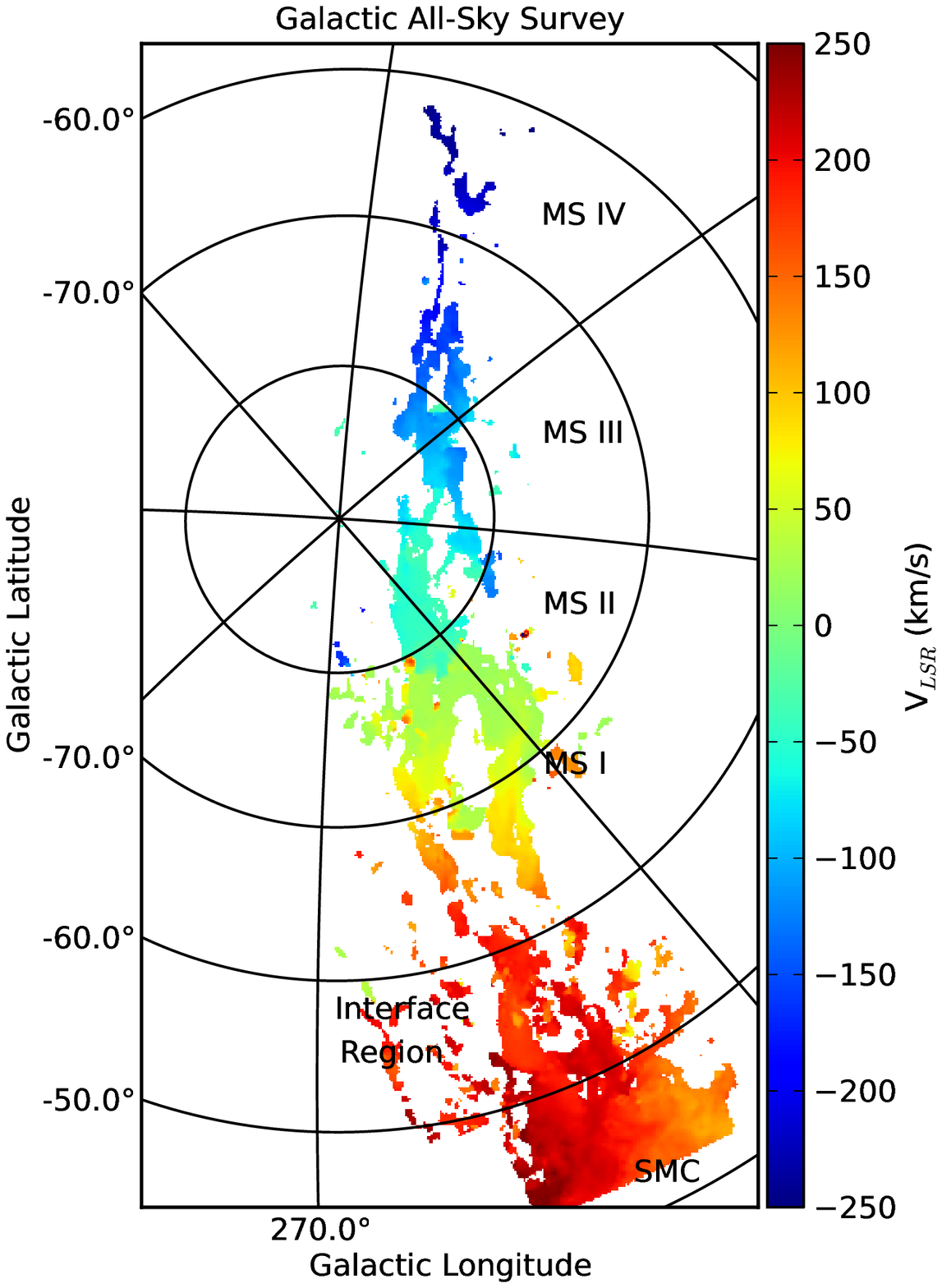}
\caption{Velocity field map in the LSR frame  
with a velocity range from $-250$ to +250~\kms\ for H75GASS (left) 
and GASS (right). The Small Magellanic Cloud, the Interface Region, 
and MS~I--IV are labeled.\label{mom1}}
\end{figure}

\begin{figure}
\epsscale{1.0}
\plottwo{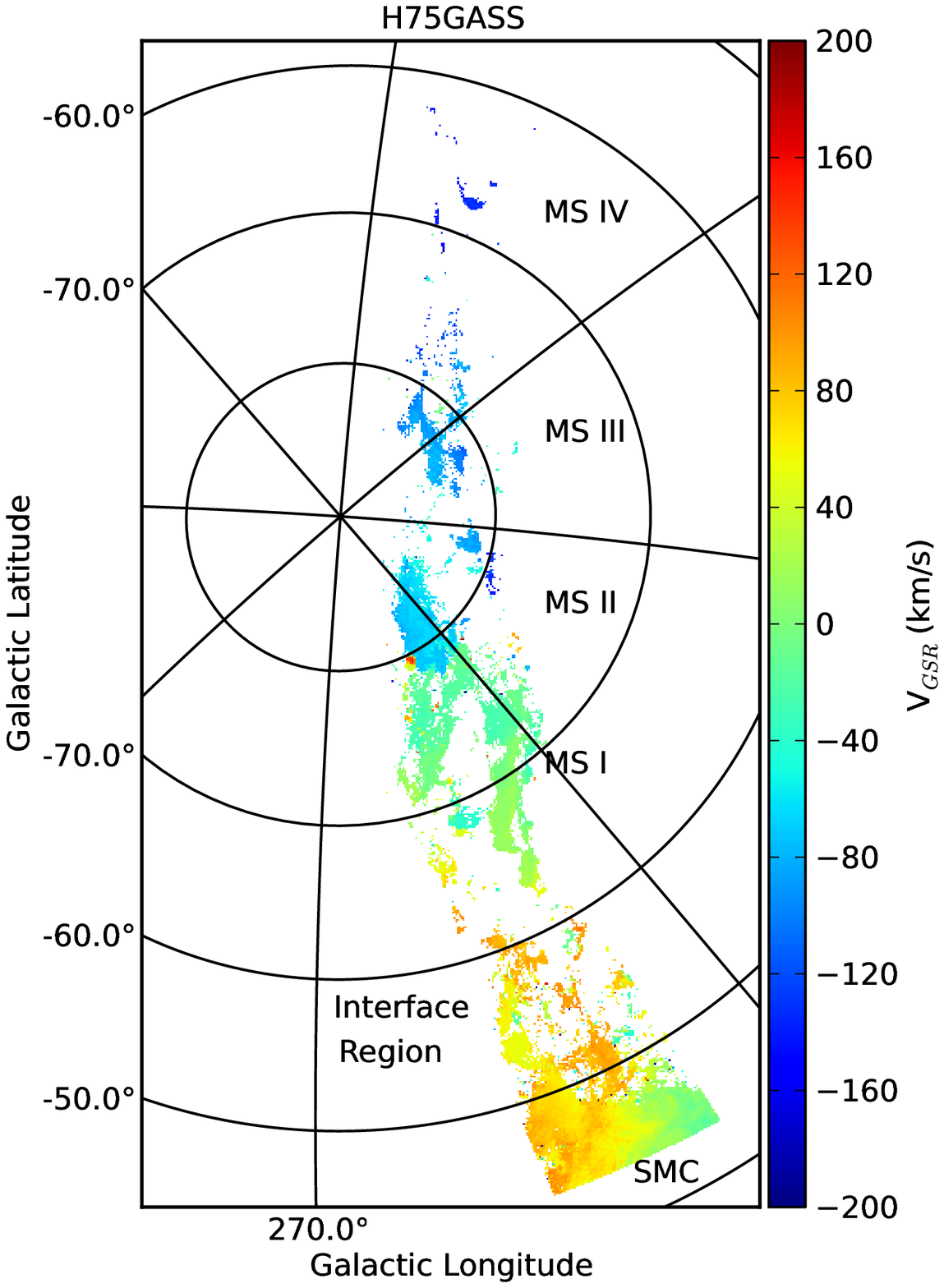}{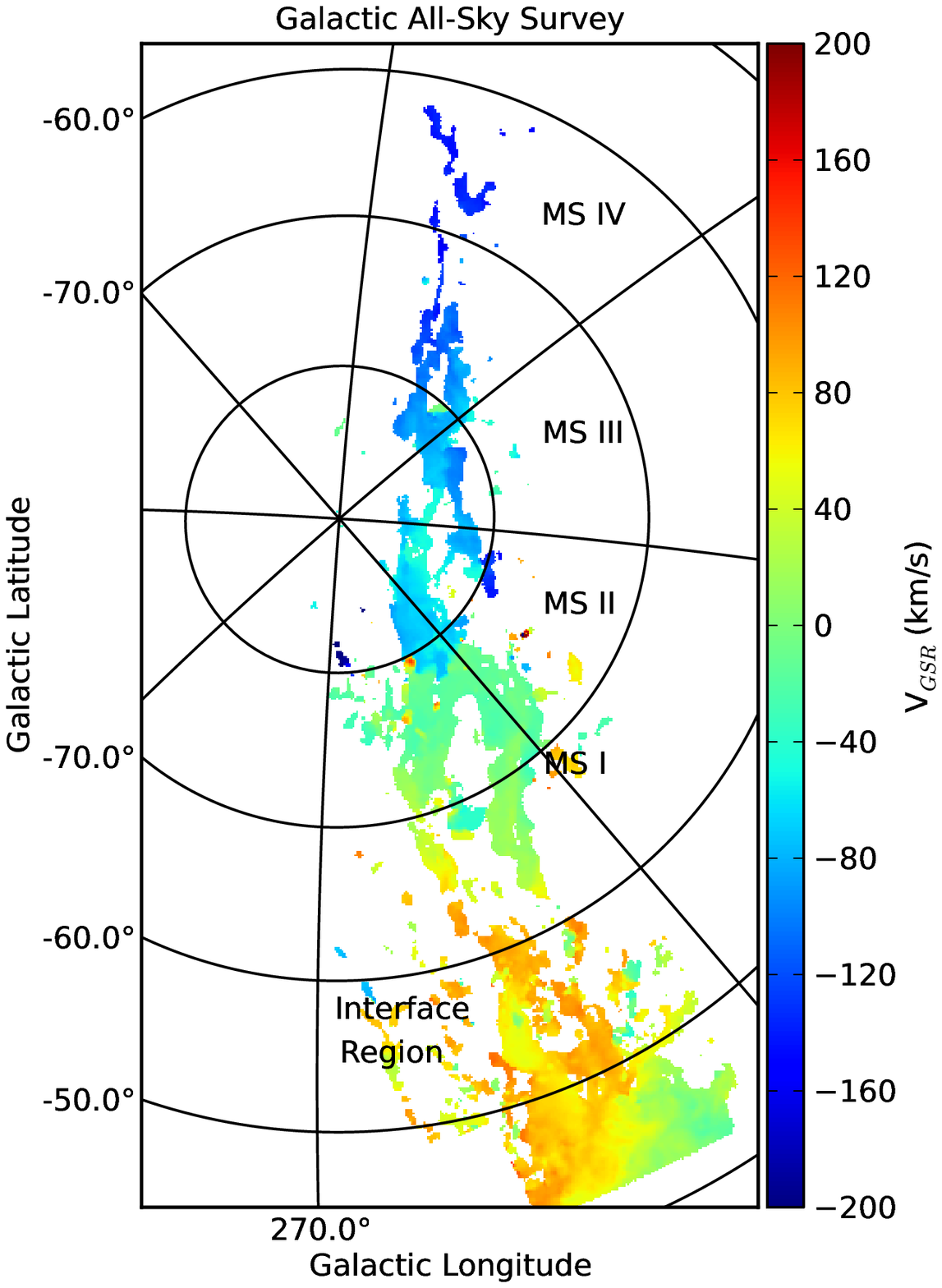}
\caption{Same as Figure~\ref{mom1}, except in GSR velocity 
reference frame with a velocity range from $-200$ to +200~\kms. 
\label{gsrmom1}}
\end{figure}

We present the velocity field of the H75GASS and GASS data in 
$V_{\rm LSR}$ in Figure~\ref{mom1} and $V_{\rm GSR}$ in Figure~\ref{gsrmom1}. 
The maps show that the observed MS covers 
a large velocity range, $-250 \lesssim V_{\rm LSR} \lesssim +250$~\kms\ and 
$-200 \lesssim V_{\rm GSR} \lesssim +200$~\kms. Faint 
and thin filamentary structures are not visible in these maps due to noise 
clipping but are visible in the integrated \HI\ column density maps 
(Figure~\ref{Hmom0} and \ref{Gmom0}).   
While it is hard to disentangle the bifurcated feature spatially in the integrated \HI\ 
column density map, the velocity field map provides a powerful tool to trace 
the filaments.  
The head of the stream starts at positive velocity and trails at 
negative velocity. The transition between the positive and negative velocity 
occurs near ($L_{\rm MS},B_{\rm MS}$) =$-50\degr$,0\degr\ (\ie, MS~II).  
This part of the stream is difficult to analyze because it extends across 
0~\kms\ and crosses in front of the Sculptor Group galaxies that have velocities 
ranging from 50 to 700~\kms\ \citep{Putman03}. Contamination 
from the Galactic emission depends on the selected velocity 
channels (see \S~\ref{prep}). 

While the velocity gradient is clearly seen along the stream, a $V_{\rm LSR}$=5.6~\kms deg$^{-1}$ 
transverse velocity gradient, in the sense that velocity decreases as a function of increasing 
declination, near $L_{\rm MS}$= $-50\degr$,   
was pointed out in \citet{Cohen82}. The exact gradient is difficult to measure 
using HIPASS due to low spectral resolution \citep{Putman03}.  
Examining the high spectral resolution of GASS and H75GASS data, 
we do not find such transverse velocity 
gradient in the specified region ($L_{\rm MS}\sim-50\degr$). However, we should point out that 
the structure in this region is contaminated by the Galactic emission. 
Proper subtraction of the Galactic emission is needed to further 
investigate the existence of the transverse velocity gradient in this particular region. 
Nevertheless, if we examine the position-velocity map in declination of the 
entire observed MS \citep{Matthews14}, a transverse velocity gradient of 
6.4~\kms deg$^{-1}$ in the GSR velocity is evident. 

\subsection{HVCs in the Magellanic Stream\label{hvcs}}
   
We have shown the on-sky distribution of HVCs found in the region of the MS 
in \S\ref{comp} (see Figure~\ref{comparisons}). 
The most noticeable pattern is the overabundance of HVCs near the South Galactic Pole 
(SGP) (right panel of Figure~\ref{comparisons}). 
HVCs in this region partially overlap in $V_{\rm LSR}$ with the Sculptor Group of 
galaxies. The coincidence between these clouds and the Sculptor Group galaxies was 
first pointed out by \citet{Mathewson75}. \citet{Putman03} carried out a comprehensive 
analysis and discussion regarding the origin of these clouds. Their conclusion is that 
these clouds are unlikely to be members of the Sculptor Group due to their velocity 
and spatial distributions not being much different from the bulk of Stream clouds.

\begin{figure}
\epsscale{0.5}
\plotone{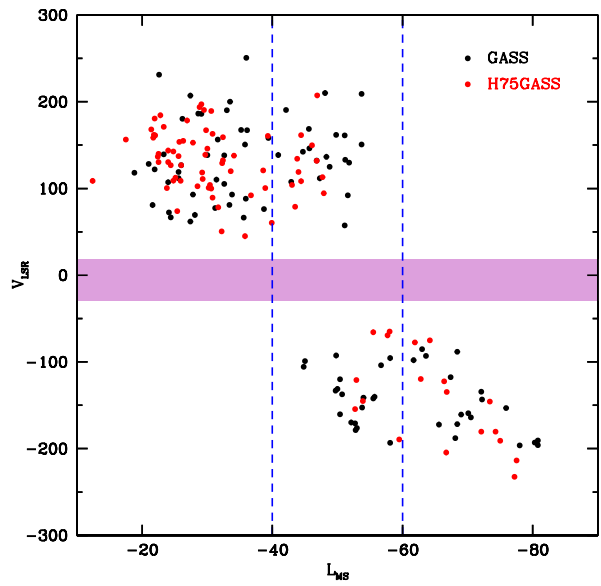}
\caption{Distribution of clouds in $L_{\rm MS}$ vs $V_{\rm LSR}$. The dotted lines mark the 
South Galactic Pole region. The black and red dots represent HVCs identified in 
GASS and H75GASS data cubes, respectively. Excluded from the plots are galaxies and HVCs 
that either overlap with the Galactic emission channels and/or lie at the spatial edge of the 
image. The shaded purple area marks the LSR velocity range from 
$-30$~\kms\ to $+18$~\kms\ that is the dominated by the Galactic gas. \label{velvslong}}
\end{figure}

Distinguishing between genuine HVCs associated with the MS and Sculptor Group is not easy based on the 
velocity information alone. Sculptor Group galaxies have a higher \HI\ 
column density. But HVCs in the MS are hard to distinguish from 
intrinsically large HVCs or tidal debris in the Sculptor Group if they exist.
We find that HVCs with large differences in velocity and \HI\ column density from the two filaments 
exist everywhere along the stream. Figure~\ref{velvslong} shows the
distribution of clouds in the $L_{\rm MS}$ -- $V_{\rm LSR}$ plane. It shows that the 
overabundance of HVCs is not unique to the South Galactic Pole. Large numbers of cloud 
are also found in the IFR ($L_{\rm MS}\sim-20\degr$ to $-30\degr$). These were not seen 
by \citet{Putman03} due to resolution and artifacts in the HIPASS data. 
Therefore, the overabundance of clouds 
in the SGP region is even less significant than before. 

To probe the interaction between the MS and the ambient halo medium, 
it is useful to study the morphology of the HVCs. 
Here we classify the HVCs in our catalog based on the morphological 
classification scheme of FSM13: 
(1) clouds with head-tail structure and velocity gradient (HT); 
(2) clouds with head-tail structure but without velocity gradient (:HT); 
(3) bow-shock shaped clouds (B); 
(4) symmetric clouds (S); and 
(5) irregular/complex clouds (IC). 

Among these groups, head-tail clouds are particularly interesting because 
their compressed head with a diffuse tail provides direct 
evidence of gas being stripped from the main condensation due to 
interaction with the surrounding ambient gas. 
Besides the traditional cometary structure, 
head-tail clouds also come in many other varieties. 
For example, they may contain additional clump of diffuse gas, or have a 
kink in the tail or velocity gradient that is 
generally associated with the \HI\ column density 
gradient. It is also known that head-tail clouds 
with a velocity gradient consist of two subgroups: those with 
the velocity of the head either leading (pHT) or lagging (nHT) tail 
(\citealp{PSM11}; FSM13). We omit head-tail clouds that do not resemble 
the obvious cometary like structure due to different viewing angles in 
this study (see \eg, \citealp{PH12}). 

\begin{figure}
\epsscale{0.8}
\plotone{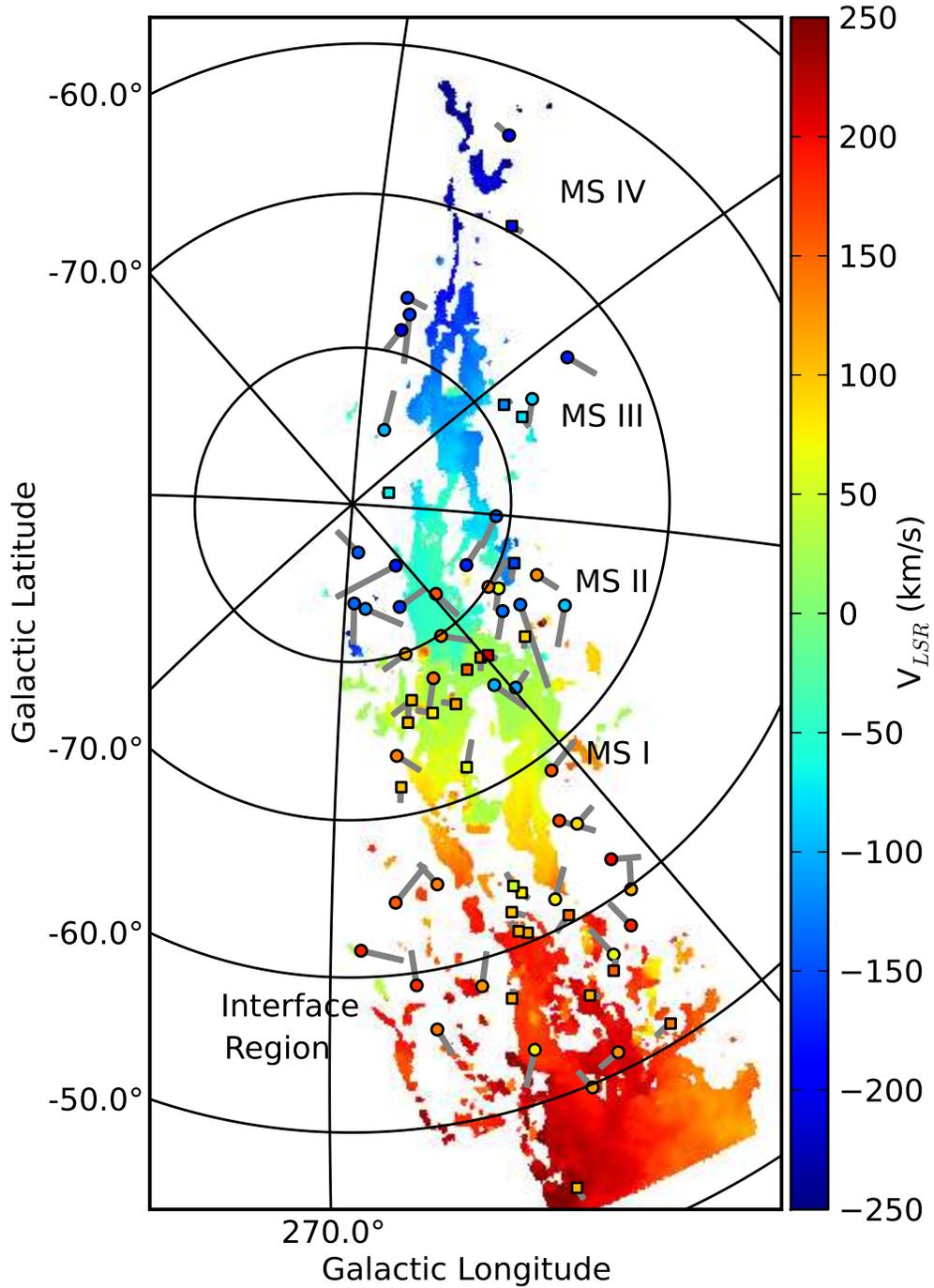}
\caption{On-sky distribution of identified head-tail clouds in 
H75GASS (squares) and GASS (circles). The colors represent the 
LSR velocity of each head-tail cloud according to the color scale 
on the right side.  
The head and tail have been enlarged from their original size 
on the plot. \label{HT_vlsr}}
\end{figure}

\begin{figure}
\epsscale{1.0}
\plottwo{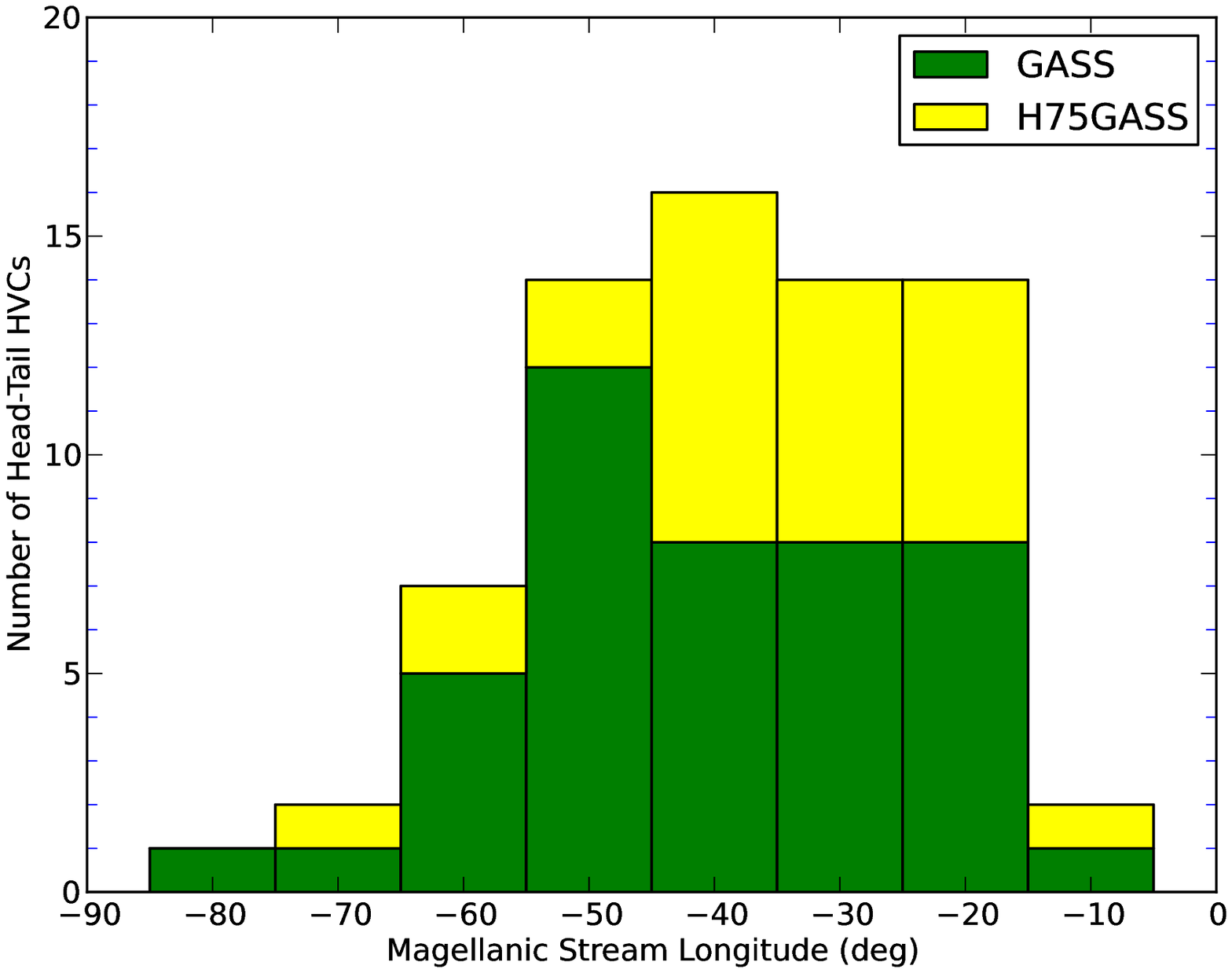}{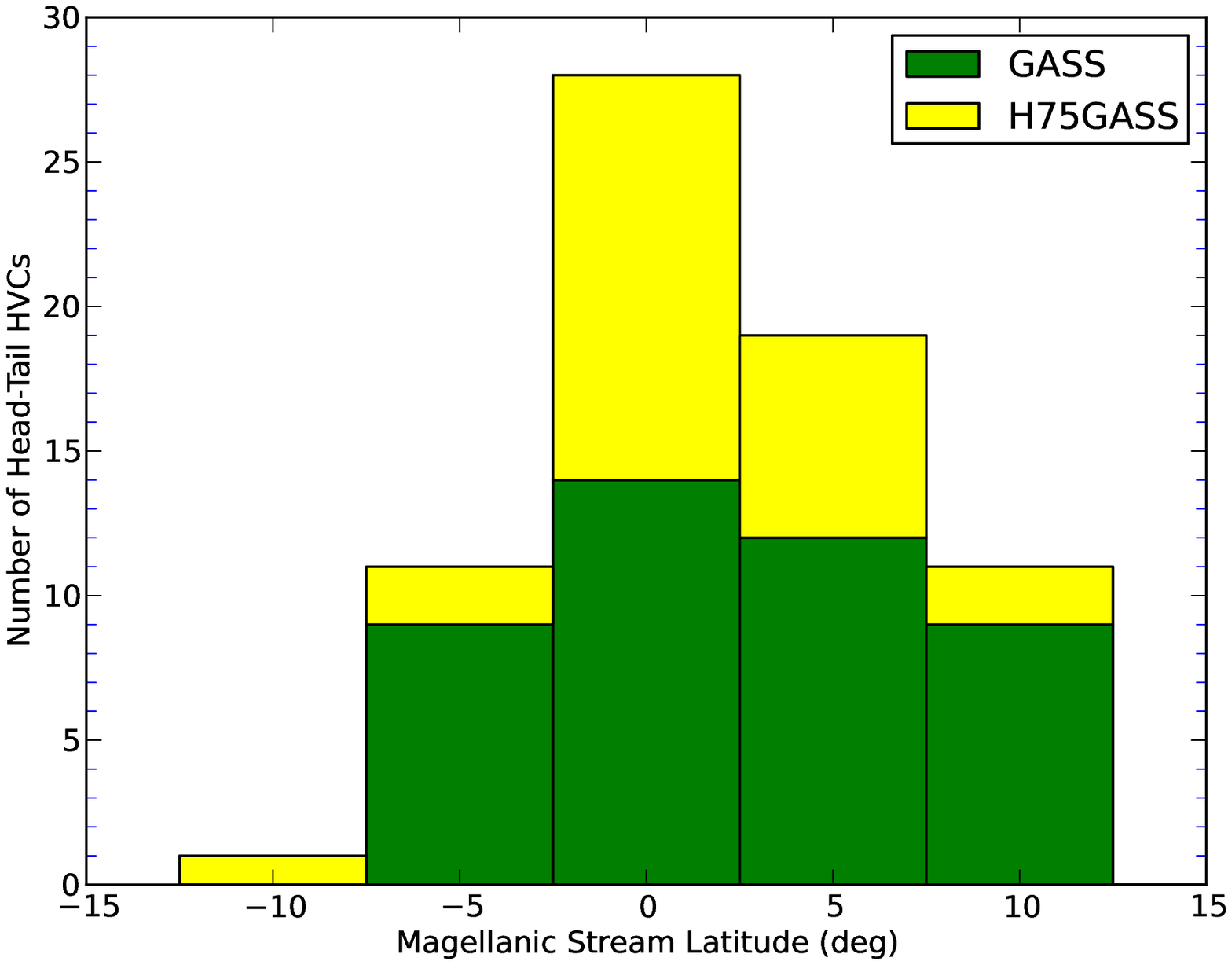}
\caption{Histograms of head-tail clouds in $L_{\rm MS}$ and $B_{\rm MS}$. 
The green and yellow represent head-tail clouds detected in GASS and 
H75GASS, respectively. 
\label{HT_mlb}}
\end{figure}

We find no obvious differences in the distributions among group 1, 2 and subgroup of pHT and nHT, 
and hence, analyze them all together here. 
In Figure~\ref{HT_vlsr}, we show the on-sky $V_{\rm LSR}$ 
distribution of head-tail clouds with and without velocity gradient 
in the region of the MS. Their distributions in $L_{\rm MS}$ and $B_{\rm MS}$ 
are shown in Figure~\ref{HT_mlb}. 
Head-tail clouds found in H75GASS data are generally compact and 
high in \HI\ column density. 
For their distribution in $V_{\rm LSR}$, the 
head-tail clouds in general follow the velocity gradient along the stream. 
Few head-tail clouds have large $V_{\rm LSR}$ difference (100--200~\kms) 
from the $V_{\rm LSR}$ of the nearby filaments. This is most obvious near the 
SGP (or MS~II) and IFR. Most head-tail clouds are at $-60\degr \lesssim 
L_{\rm MS} \lesssim -20\degr$ and $-5\degr \lesssim B_{\rm MS} \lesssim +10\degr$. 
Compared to the distribution of all HVCs in $B_{\rm MS}$ (Figure~\ref{dist}), 
most of the clouds at $B_{\rm MS}\sim$10\degr\ are head-tail clouds. 

\begin{figure}
\epsscale{1.0}
\plotone{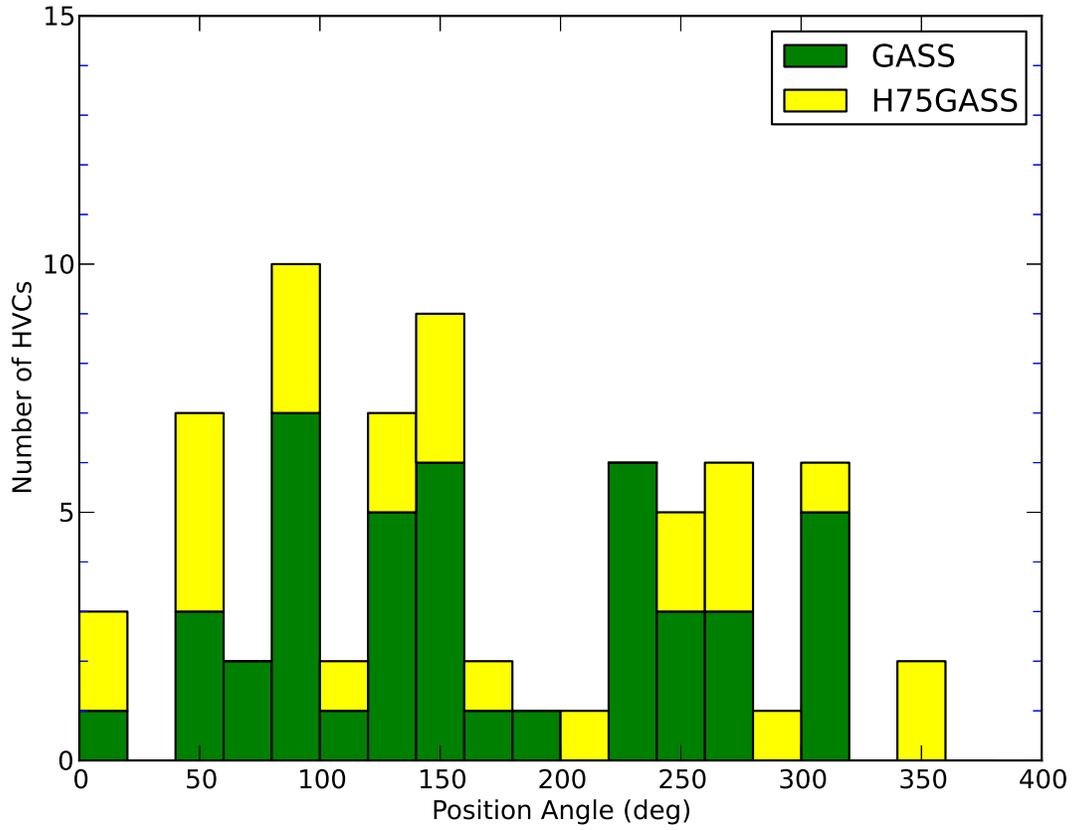}
\caption{Histogram of position angles relative to the MS coordinates for 
all the head-tail clouds. 0\degr\ is parallel to the $B_{\rm MS}$ and the angle turns 
anti-clockwise. 90\degr\ infers the head of the cloud is pointed away 
from the general motion of the Magellanic System and parallel to the $L_{\rm MS}$. 
The green and yellow represent head-tail clouds detected in GASS and 
H75GASS, respectively. \label{HT_PA}}
\end{figure}

The pointing direction of the head-tail clouds is also shown in Figure~\ref{HT_vlsr}, 
with head and tail being enlarged for better visibility. 
Due to the measurement errors in position angle, in some cases 
we manually inspected the  cloud and adjusted the values 
whenever necessary to better represent the pointing direction 
of the clouds (as discussed in FSH13).   
In Figure~\ref{HT_PA}, we show the histogram of the 
pointing direction of the head-tail clouds. There is no preferential 
pointing direction but careful examination reveals a nearly equal number pointing 
either toward (225\degr--315\degr) or away (45\degr--135\degr) from the head of the stream.
We also find a total number of 70 head-tail clouds, which is $\sim28\%$ of all HVCs 
in our catalog. 60$\%$ (42/70) of the head-tail clouds 
show a clear velocity gradient, with a nearly equal number of clouds 
for the velocity gradient subgroups, \ie, 20 vs 21 for nHT and pHT, respectively. 
The percentages are very similar to those found in the LA region 
(FSM13). We have about a factor of six more head-tail clouds in this study 
than the study by \citet{PSM11} in the same region. This may partly be due to 
different selection criteria.
While the LA region still has a factor of 1.5 more head-tail clouds than the MS 
region, this finding is unexpected because disturbed clouds are absent in MS~VI 
\citep{Stanimirovic08}. 
We will discuss the implication of the pointing direction and large number of head-tail 
clouds in the MS in \S~\ref{diss}. 

\begin{figure}
\epsscale{1.0}
\plottwo{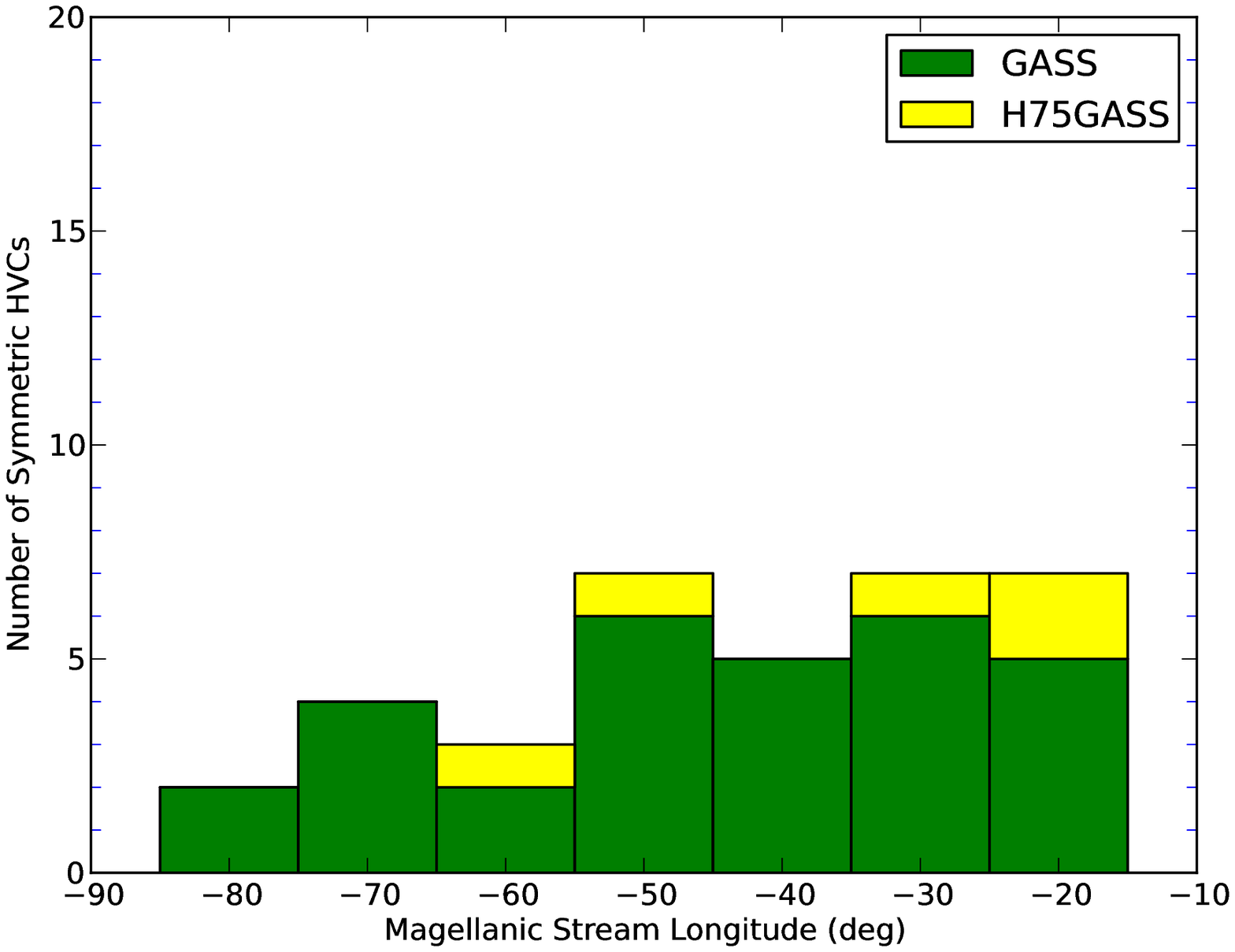}{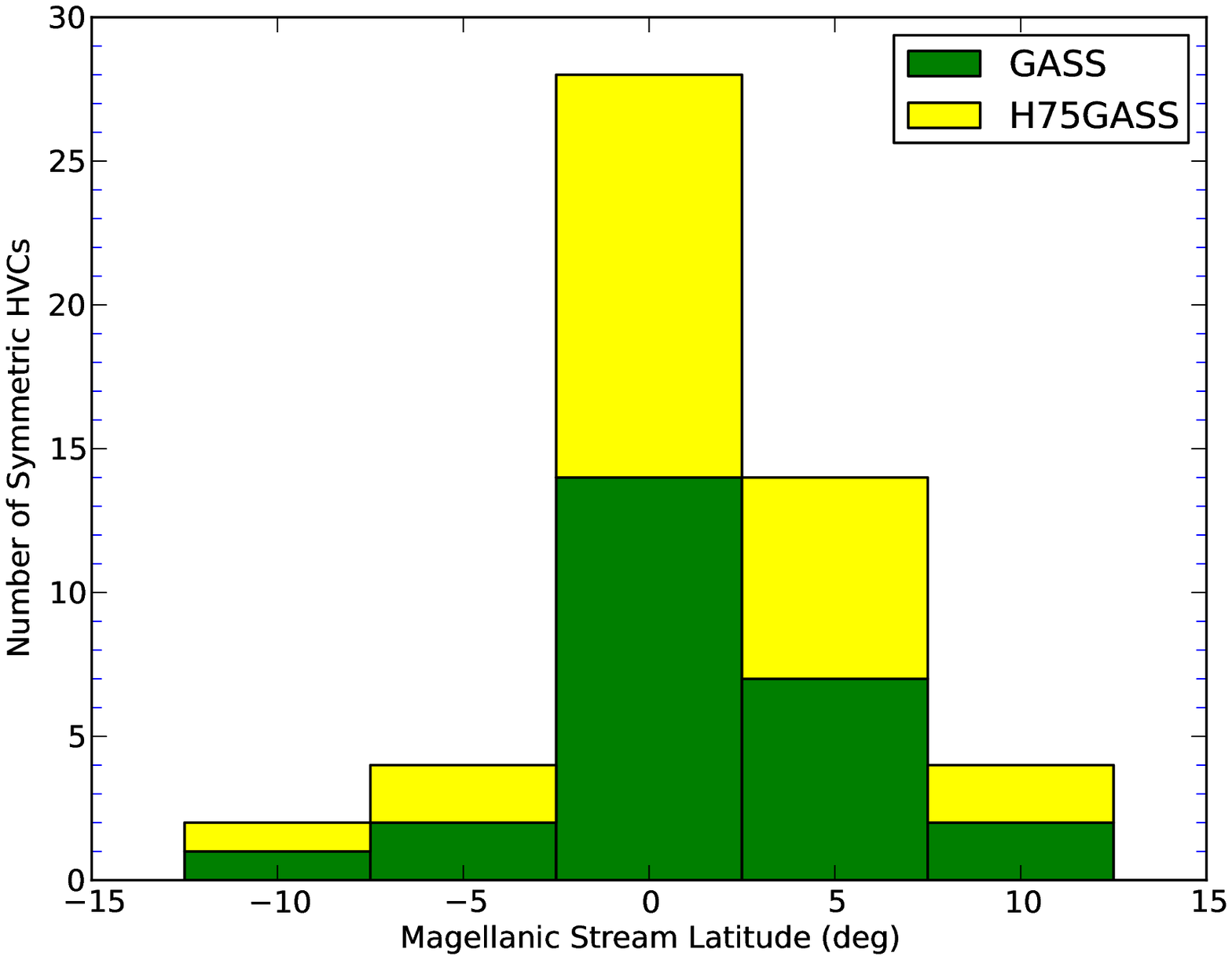}
\caption{Same as Figure~\ref{HT_mlb}, but for symmetric clouds. 
\label{sy_mlb}}
\end{figure}

\begin{figure}
\epsscale{0.6}
\plotone{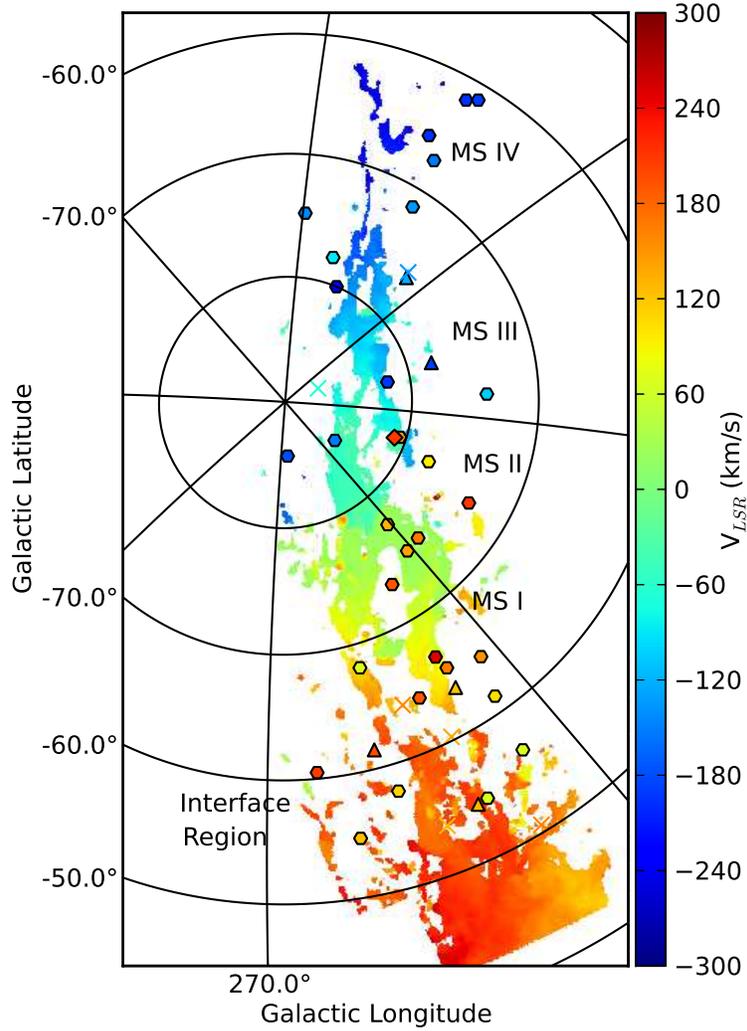}
\caption{On-sky distribution of symmetric and bow-shock shaped clouds. 
The colors represent the 
LSR velocity of each cloud 
according to the color scale on the right side. The diamonds, 
and crosses represent the bow-shock-shaped clouds for the 
GASS and H75GASS data, respectively. 
The hexagons and triangles represent the symmetric clouds for 
the GASS and H75GASS data, respectively. 
\label{SB_dist}}
\end{figure}

The bow-shock-shaped clouds have a similar morphological structure 
to the head-tail clouds. They are characterized by a compressed head but with two 
deflected gas wings instead of a tail. Many studies suggest that 
bow-shock-shaped cloud are less common than head-tail clouds despite both of 
them showing evidence of ram-pressure interaction with the ambient medium 
(see \eg, \citealp{Westmeier05}; FSM13). 
On the other hand, symmetric clouds morphologically do not show signs of disturbance, 
although head-tail cloud pointed directly along the line of sight can appear as a symmetric cloud.
The total numbers of bow-shock-shaped and symmetric clouds are 8 and 35 
in our catalog, respectively.  
This is about half the number of symmetric clouds in the MS as compared 
to the LA region.  
Figure~\ref{SB_dist} shows the distributions of bow-shock-shaped (diamonds 
and crosses) and symmetric clouds (hexagons and triangles) 
in $V_{\rm LSR}$. Distributions of symmetric clouds in $L_{\rm MS}$ and $B_{\rm MS}$ 
are shown in Figure~\ref{sy_mlb}. They are distributed evenly 
across the entire range of $L_{\rm MS}$, 
but mainly populate the range of $\sim0\degr$--5\degr\ in $B_{\rm MS}$. 
The $V_{\rm LSR}$ distribution is quite similar to the one for head-tail clouds. 
Velocity gradients are common among the symmetric clouds. 
One notable, distinctive bow-shock-shaped cloud is HVC+73.7$-$67.5$-213$, which is 
located in MS~IV. It is a relatively large (1.2\degr) 
compared to others found in this study. 

\section{Discussion\label{diss}}

The formation of the MS has previously been attributed to either tidal or 
ram-pressure mechanisms. 
Many efforts have been put into creating tidal models that 
accurately reproduce the global morphology such as 
the bifurcation of the MS. 
However, these tidal models fail to reproduce some of the 
observed features, such as the gradual decrease in \HI\ column density 
along the stream. 
Ram-pressure models, on the other hand, are able to help overcome this problem. 
Thus, the formation of the MS might be due to 
a combination of ram-pressure and tidal stripping. 
Implementing both formation mechanisms simultaneously into 
simulations is not an easy task due to the  
large number of particles required to resolve the ISM and hot gas interaction.
A notable attempt to employ both physical mechanisms is 
the study of \citet{Mastropietro05}. The drawback of their model is 
the sole consideration of 2-body tidal interaction (MW-LMC), whereas in many  
tidal models, the SMC plays an important role 
in the formation of the MS. 
Metallicity measurements 
of $FUSE$ and $HST$ spectra suggest that most of the MS 
originates from the SMC \citep{Sembach01,Fox10,Fox13}. 
In this section, we mainly discuss the effect of ram-pressure stripping on the MS. 

Aspects such as overall morphology and decrease in \HI\ column density of the two 
filaments can be explained by the ram-pressure model and evaporation of gas clouds. 
Models of ram pressure stripped galaxies (viewed edge-on) 
have shown that ram-pressure is sufficient to remove part of the 
neutral gas from the disk \citep{Mastropietro05}. 
The stripped gas trails behind the galaxy, and 
the morphology at early times looks like a bow shock  
(see \eg, Figure~4 of \citealp{Mastropietro05}). 
As the interaction time continues, the gas in the disk decreases significantly. 
Eventually, islands of complex filaments are formed in the 
trailing direction and fan out near the end \citep{Steinhauser12}, 
which is morphologically similar to the MS that we see today \citep{Stanimirovic08}.
Clouds that are stripped first would have longer interaction time with the hot halo 
and sustain larger loss of gas due to evaporation.

To further investigate the effect of ram-pressure on a smaller scale, we 
examine the distributions of head-tail clouds and their pointing direction. 
These head-tail clouds are presumably the debris of larger filaments 
broken off due to ram-pressure stripping. 
Thus, they are expected to follow the motion of the larger filaments, \ie, 
with their head pointing toward the SMC and their tail pointing 
in the direction of decreasing 
$L_{\rm MS}$.  
However, they seem to be pointing in a random direction, 
similar to the finding in the LA region (FSM13). 
This is a rather surprising result because the 
random motion of head-tail clouds in the region of the LA 
is presumably caused by strong turbulence generated from the collision between   
the LA and the Galactic disk gas. This scenario of incoming 
warm neutral gas colliding with the hotter ambient medium  
\citep{AH05} is inadequate to explain the random pointing 
of head-tail clouds in the MS region, even though turbulence 
must be at play as well. We look into the cascade scenario, 
which was originally proposed to explain the 
observed H$\alpha$ emission along the MS \citep{BH07}. 
In the hydrodynamical simulation of \citet{BH07}, the upstream clouds 
experience gas depletion via Kelvin-Helmholtz (KH) instability due to 
interaction with the hot halo. The depleted gas plows 
into the following cloud, which leads to shock ionization and depletion 
of the downstream clouds. The entire process continues 
downstream like a chain-reaction. During this process, 
the downstream clouds transfer momentum to the depleted gas in the front, 
which results in Rayleigh-Taylor (RT) instabilities. 
With both KH and RT instabilities, the process becomes highly 
turbulent for the entire region. 
This scenario agrees with the observational properties of 
head-tail clouds in the MS.

The life time of the disrupted clouds in the cascade scenario is 100--200~Myr, which is 
relatively short as compared to the age of the stream (1--2 Gyr) 
as estimated from simulations. 
Constant replenishment of gas from the MCs is needed to 
reproduce the observed structure \citep{BH07}. 
Other physical parameters can also 
lengthen the lifetime and govern the stability 
of clouds, for example, magnetic field, dark-matter, heat conduction, 
size and, halo and cloud densities. 
Two and/or three-dimensional hydrodynamical simulations 
have experimented with these parameters 
(see \eg, \citealp{QM01,PH12,HP09,VH07}).

In \citet{QM01}, the behavior of pure gas clouds and dark-matter dominated HVCs interacting with the 
different halo environment was investigated. 
They found that head-tail structures emerge when the 
ambient environment reaches a density of 10$^{-4}$~cm$^{-3}$ with a cloud 
velocity of 200~\kms. HVCs with a dark-matter component survive longer ($\sim10$~Gyr) 
as compared to those without ($\sim10$~Myr). However, the 
same cloud was not tested with and without dark-matter.
The recent \citet{PH12} models considered 
this effect and found that dark-matter free HVCs can easily survive 
for 100~Myr. Their model also suggests that dark-matter free clouds 
are physically much similar to the observed large HVC complexes 
with a heterogeneous 
multi-phase internal gas structure. 

The effect of heat conduction for stabilizing the HVCs (or Giant Molecular Clouds) 
is explored by \citet{VH07}. 
Their models show that clouds with two-phase structure can stabilize and 
survive longer as a result of 
heat conduction. 
While heat conduction may explain the longer survival timescale of 
head-tail clouds in the MS, massive clouds like the filaments of the MS 
are yet to be tested in simulations (G. Hensler, priv. comm.). 
\citet{HP09} also explored effects of various physical parameters 
based on two experimental setups: wind-tunnel and free-fall. 
Their results show that clouds 
with \HI\ masses less than 10$^{4.5}$~$M_{\sun}$ will lose their 
gas after moving only 10~kpc or less under typical halo density and 
relative velocity conditions. The visibility of the head-tail structure 
depends on the viewing angle. 
Nevertheless, all the hydrodynamical simulations have so far only considered 
clouds at lower $z$ (distance from the Galactic plane) 
due to the large uncertainty in the halo density beyond 10~kpc. 
This implies that these models might not be suitable for explaining the evolution of 
HVCs originating from the MCs. Interestingly, HVCs created with these models 
are morphologically and physically similar to what we observe, \eg, the 
bow-shock-shaped cloud, HVC+73.7$-$67.5$-213$, and the head-tail clouds.

There is an overall larger number of head-tail clouds in the MS region 
found in this study as compared to the study by \citet{PSM11} 
(see \S\ref{hvcs}). 
It is possible that we missed some of the head-tail clouds  
not resembling classical head-tail morphologies due to different  
viewing angles (see \citealp{PH12}), which will increase the number of head-tail clouds.
While a larger number of head-tail clouds is found, there is a significant 
decrease in number toward the tail end 
($L_{MS} >-60\degr$; see the left panel of Figure~\ref{HT_mlb}). 
This is consistent with the 
finding of a lack of head-tail clouds in the region of MS~IV \citep{Stanimirovic08}, 
which suggests that ram-pressure stripping has less of an effect on the tail end of the stream. 
This is not surprising given that the predicted model distance is the largest at the tail end 
of the stream, which the ram-pressure decreases as a function of distance square 
in an assumed isothermal halo model. It is also possible that 
the tail of the MS is aligned radially along the line-of-sight 
(see Figure~4 of \citealp{DB12}), in which case viewing geometry 
would reveal more symmetric cloud shapes.
 
We find that the number of symmetric clouds is about a factor of two 
smaller than the number of head-tail clouds in the range 
$-60\degr < L_{\rm MS} < -20\degr$. A slight increase in the number of symmetric clouds 
as compared to head-tail clouds is observed beyond $L_{\rm MS}\sim-60\degr$. 
We know that the MS~VI region is exclusively populated with 
symmetric clouds \citep{Stanimirovic08}. This raises the question whether 
we will see a continual trend of increasing number of symmetric clouds 
in the MS~V region, which is 
outside our currently studied area. If the answer is yes, then this would 
further confirm the 
model prediction of distance 
increasing as a function of $L_{\rm MS}$. 
With this in mind, it is rather puzzling to see 
that the northern extension of the stream \citep{Nidever10} has a very different morphology 
from other parts of the MS. 
It morphologically looks more like the LA (FSM13). 
A statistical analysis of different morphological types 
of HVCs in the northern extension might shed some light on 
the distance scale of the farthest end of the stream.

\section{Summary and Conclusions\label{concl}}

We have compiled a catalog of HVCs in the Magellanic Stream using 
new ATCA and Parkes (GASS) data. The catalog includes 251 HVCs and 2 galaxies. 
Most of the clouds detected in the combined H75GASS are unique. 
We used $Duchamp$ as a source finding tool 
and to parametrize the detected sources. Most of the parameters were derived 
from $Duchamp$ except the angular size and the peak \HI\ column density, 
which were determined independently. We made a comparison 
between our catalog and \citet{Putman02}, who employed the reprocessed 
HIPASS data for their study. 

We have presented the distributions of 
clouds and their properties. 
The HVCs agglomerate in a narrow Magellanic Stream latitude ($B_{\rm MS}$) range of $\pm10\degr$ and 
decrease linearly with decreasing Magellanic Stream longitude ($L_{\rm MS}$). 
The kinematics show bimodality in $V_{\rm LSR}$ 
but none in $V_{\rm GSR}$. The majority of the clouds detected in 
GASS have lower \HI\ column density than in H75GASS. The overall 
velocity FWHM distribution is similar to that found in the Leading Arm region (FSM13). 

The overall morphology and kinematics of the MS have been revisited. We find 
that the two filaments of the MS are extended into the IFR, connected to the SMC 
and extend into the northern end of the Magellanic Bridge. 
The former finding is consistent with the study by \citet{Nidever08}, but 
differs from the conclusion of \citet{Bruns05}. The connection to the SMC and 
Magellanic Bridge is 
contradictory to the conclusions of \citet{Nidever08}.
Our suggestion that the 
two filaments of the MS are twisted and largely emanate from the SMC 
is consistent with the \citet{Fox13} metallicity measurements, 
which show 0.1 solar metallicity along much of the Stream.
While we could not 
investigate the existence of the transverse velocity gradient as seen in 
\citet{Cohen82} in this study, we found 
a clear transverse velocity gradient of 6.4~\kms deg$^{-1}$ in the GSR 
velocity frame based on the 
position-velocity map of \citet{Matthews14}. 

We find that the MS has a complex filamentary structure.  
Morphological classification is presented and distributions 
of each type are discussed. In contrast to the studies by \citet{Stanimirovic08} 
and \citet{PSM11}, we found a large number of head-tail clouds in 
the MS region. This strongly suggests that ram-pressure stripping plays an important role 
in the formation mechanism of the MS. 
The pointing direction of the head-tail 
clouds appear to be random, which suggests the presence of strong turbulence. 
The turbulence induced by the cascade scenario 
is a likely cause. We also discussed various physical parameters 
that can lengthen the lifetime of the HVCs. While hydrodynamical 
simulations of HVC may not explain the HVCs origin from 
the MCs, they do show how small scale structure may arise. 
Evaporation of gas clouds is 
also responsible for the decrease of \HI\ column density as seen toward the tail end 
of the MS. 

\acknowledgments

BQF is the recipient of a John Stocker Postdoctoral Fellowship from the 
Science and Industry Research Fund. This research made use of APLpy, 
an open-source plotting package for Python hosted at 
http://aplpy.github.com. 
This publication also made use of data products from Parkes and  
Narrabri radio telescopes. The  
Australia Telescope Compact Array/Parkes radio telescope is 
part of the Australia Telescope National Facility, 
which is funded by the Commonwealth of Australia 
for operation as a National Facility managed by CSIRO. 
We thank Kenji Bekki and Cameron Yozin for generating fruitful discussion, 
Tobias Westmeier for providing his parametrization code and helpful comments 
and Matthew Whiting for assisting with the running of $Duchamp$ on 
large data cubes.

\bibliographystyle{apj}
\bibliography{ref}

\clearpage

\clearpage
\input{tables}

\end{document}

%% file: tables.tex
\begin{deluxetable}{ccc}
\tablewidth{0pt}
\tablecolumns{3}
\tablecaption{The velocity ranges of the original cubes and 
extracted subcubes.\label{cubesvel}}
\tablehead{
\colhead{Cube} & 
\colhead{Full $V_{\rm LSR}$ range} &
\colhead{Extracted $V_{\rm LSR}$ range}\\
\colhead{} &
\colhead{(\kms)} &
\colhead{(\kms)} 
}
\startdata
1	&	$-111.45\ldots+382.05$	&	$+45.17\ldots+279.58$	\\
2	&	$-111.21\ldots+382.28$	&	$+23.97\ldots+261.65$	\\
3	&	$-110.55\ldots+382.95$	&	$+18.03\ldots+214.42$	\\
4	&	$-109.61\ldots+383.89$	&	$+18.97\ldots+190.58$	\\
5	&	$-324.38\ldots+330.30$  &       $+18.90\ldots+227.87$	\\
	&	                 	&	$-110.30\ldots-31.19$	\\
6	&	$-323.07\ldots+331.60$	&	$-201.25\ldots-31.54$	\\
7	&	$-320.97\ldots+333.70$	&	$-209.04\ldots-31.09$	\\
8	&	$-318.94\ldots+335.75$	&	$-249.81\ldots-48.84$	\\
9	&	$-317.83\ldots+338.88$	&	$-255.28\ldots-72.45$	\\
10	&	$-315.68\ldots+339.02$	&	$-259.72\ldots-114.79$	\\
11	&	$-314.30\ldots+340.40$	&	$-255.05\ldots-149.68$	
\enddata
\end{deluxetable}

\begin{deluxetable}{cccccccccccccccccc}
\rotate
\tabletypesize{\tiny}
\tablewidth{0pt}
\tablecolumns{18}
\tablecaption{Catalog of detected sources in the region 
of the Magellanic Stream.\label{catalog}}
\tablehead{
\colhead{ID} & 
\colhead{Designation} &
\colhead{$L_{\rm MS}$} &
\colhead{$B_{\rm MS}$} &
\colhead{$V_{LSR}$} &
\colhead{$V_{GSR}$} &
\colhead{$V_{LGSR}$} &
\colhead{FWHM} &
\colhead{$F_{\rm int}$\tablenotemark{a}} &
\colhead{$T_{\rm B}$} &
\colhead{$N_{\rm HI}$}  &
\colhead{Semi-major}  & 
\colhead{Semi-minor}  & 
\colhead{PA}  &
\colhead{Flag\tablenotemark{b}} & 
\colhead{Classification\tablenotemark{c}} &
\colhead{Data\tablenotemark{d}}  &
\colhead{Comment}  \\
\colhead{} & 
\colhead{($l \pm b +V_{LSR}$)} &
\colhead{$\degr$} &
\colhead{$\degr$} &
\colhead{\kms} &
\colhead{\kms} &
\colhead{\kms} &
\colhead{\kms} &
\colhead{Jy km s$^{-1}$} &
\colhead{K} &
\colhead{10$^{19}$ cm$^{-2}$}  &
\colhead{$\degr$}  & 
\colhead{$\degr$}  & 
\colhead{$\degr$}  &
\colhead{} & 
\colhead{} & 
\colhead{} & 
\colhead{}  \\
\colhead{(1)} & 
\colhead{(2)} &
\colhead{(3)} &
\colhead{(4)} &
\colhead{(5)} &
\colhead{(6)} &
\colhead{(7)} &
\colhead{(8)}  &
\colhead{(9)}  & 
\colhead{(10)}  & 
\colhead{(11)}  &
\colhead{(12)} & 
\colhead{(13)} &
\colhead{(14)} &
\colhead{(15)} &
\colhead{(16)} &
\colhead{(17)} &
\colhead{(18)} \\
}
\startdata
24	&	HVC+347.9$-$83.8$-$192	&	$-$55.0	&	+01.7	&	\nodata	&	\nodata	&	\nodata	&	\nodata	&	\nodata	&	\nodata	&	\nodata	&	\nodata	&	\nodata	&	\nodata	&	SR	&	\nodata	&	G	&		\\
25	&	HVC+072.3$-$70.0$-$191	&	$-$75.0	&	$-$01.5	&	$-$191.0	&	$-$119.3	&	$-$079.8	&	21.1	&	12.2	&	1.32	&	2.84	&	0.4	&	0.2	&	17	&	\nodata	&	IC	&	H	&		\\
26	&	HVC+064.3$-$61.3$-$191	&	$-$80.8	&	$-$08.8	&	$-$190.6	&	$-$095.5	&	$-$060.4	&	21.0	&	3.5	&	0.23	&	0.36	&	0.2	&	0.2	&	45	&	\nodata	&	S	&	G	&		\\
27	&	HVC+019.8$-$78.1$-$189	&	$-$59.5	&	$-$04.2	&	$-$189.4	&	$-$174.0	&	$-$149.0	&	5.4	&	1.5	&	1.36	&	1.18	&	0.1	&	0.1	&	$-$67	&	\nodata	&	S	&	H	&		\\
28	&	HVC+079.6$-$78.5$-$188	&	$-$68.1	&	+03.7	&	$-$187.9	&	$-$144.5	&	$-$104.6	&	13.7	&	5.2	&	0.27	&	0.37	&	0.3	&	0.3	&	$-$76	&	\nodata	&	pHT	&	G	&		\\
29	&	HVC+069.2$-$72.8$-$180	&	$-$72.1	&	$-$01.1	&	$-$180.4	&	$-$119.6	&	$-$081.6	&	45.0	&	27.1	&	1.40	&	3.95	&	0.8	&	0.3	&	$-$15	&	\nodata	&	IC	&	H	&		\\
30	&	HVC+066.3$-$69.6$-$180	&	$-$74.3	&	$-$03.5	&	$-$180.3	&	$-$110.1	&	$-$073.2	&	22.5	&	9.2	&	1.57	&	3.52	&	0.1	&	0.1	&	78	&	\nodata	&	:HT	&	H	&		
\enddata
\tablecomments{Table~\ref{catalog} is published in its entirety in the
electronic edition of the {\it Astrophysical Journal}.  
A portion is shown here for guidance regarding its form and content.}
\tablenotetext{a}{Corrected $F_{\rm int}$.}
\tablenotetext{b}{SR: the detection lies at the edge of the spectral region; 
E: the detection is next to the spatial edge of the image.}
\tablenotetext{c}{HT: head-tail cloud with velocity gradient; :HT: head-tail 
cloud without velocity gradient; S: symmetric cloud; B: bow-shock cloud; 
IC: irregular/complex cloud.}
\tablenotetext{d}{H: H75GASS; G: GASS.}
\end{deluxetable}